\documentclass[12pt,a4paper]{article}
\usepackage[dvips]{lscape,graphicx}
\usepackage{axodraw}
\usepackage[cp1251]{inputenc}
\usepackage[russian]{babel}

\newcommand{\bc}{\begin{center}}
\newcommand{\ec}{\end{center}}
\newcommand{\bd}{\begin{displaymath}}
\newcommand{\ed}{\end{displaymath}}
\newcommand{\be}{\begin{equation}}
\newcommand{\ee}{\end{equation}}
\newcommand{\ba}{\begin{array}}
\newcommand{\ea}{\end{array}}
\newcommand{\bea}{\begin{eqnarray}}
\newcommand{\eea}{\end{eqnarray}}
\newcommand{\bt}{\begin{tabular}}
\newcommand{\et}{\end{tabular}}

\newcommand{\bp}{\begin{picture}}
\newcommand{\ep}{\end{picture}}
\newcommand{\bfi}{\begin{figure}}

\begin{document}
\thispagestyle{empty}
\begin{flushright}ITEP 03-01
\end{flushright}
\thispagestyle{empty}

\vspace*{1cm}

\vspace{5mm}

\parbox[b]{150mm}{\bc \Large\bfseries Selected Theoretical Issues in $B$ Meson Physics:\\ CKM
matrix and  Semileptonic Decays
 \ec}
 \vspace{2cm}

 \bc {\large I.M.Narodetskii\\[4mm]
 Institute of Theoretical and Experimental Physics, Moscow}

\ec

\vspace{3cm} \bc {\bf\Large Abstract}\ec \vspace{1cm}

\noindent These notes are a written version of a lecture given at
the International Seminar {\it Modern Trends and Classical
Approach} devoted to the 80$^{th}$ anniversary of Prof. Karen
Ter-Martirosyan, ITEP September 30~-~October 1, 2002. The notes
represent a non-technical review of our present knowledge on the
phenomenology of weak decays of quarks, and their r\^ole in the
determination of the parameters of the Standard Model. They are
meant as an introduction to some of the latest results and
applications in the field.  Specifically, we focus on $CP$
violation in $B$-decays and the determination of the CKM matrix
element $V_{cb}$ from semileptonic decays of $B$ mesons. We also
briefly discuss phenomenological applications concerning the
electron-energy spectra in semileptonic $B$ and $B_c$ decays.

\newpage

\setcounter{page}{1}
\renewcommand{\contentsname}{Contents}
\tableofcontents

\newpage


\vspace*{3mm}

\par

\section{Introduction}
The goal of the $B$ physics is to precisely test the flavor
structure of the Standard Model (SM) {\it i.e.} the
Cabibbo-Kobayashi-Maskawa (CKM) \cite{CKM} description of quark
mixing and $CP$ violation. Flavor physics played an important role
in the development of the SM. For a long time the  only
experimental evidence for $CP$ violation came from the Kaon
sector: $|\epsilon_K|=(2.280\pm0.013)\times 10^{-3}$ \cite{pdg},
$\epsilon_K/\epsilon'_K=(1.66\pm0.16)\times
10^{-3}$~~\footnote{The world average based on the recent results
from NA48 and KTeV experiments and previous results from NA31 and
E731 collaborations, quoted from \cite{Buras03}} . The smallness
of $K^0-\bar K^0$ mixing led to the GIM compensation mechanism and
a calculation of a $c$ quark mass before it has been discovered
\cite{JIM70}. The existence of $CP$ violation in neutral kaon
decay provoked the hypothesis of a third generation, four years
before  experimental discovery of the $\Upsilon$ particles, the
first experimental detection of the $b$ quark. Surprising
discovery of the large $B^0-\bar B^0$ mixing \cite{Argus1997} was
the first evidence for a very large top quark mass. The
implications of this observation were important for the
experimental program on CP violation. Two high luminosity B
factories  (SLAC/BaBar and KEKB/Belle) were commissioned with
remarkable speed in late 1998. The experiments starting physics
data taking 1999. In the summer of 2001, BaBar and Belle
experiments announced the observation of the first statistically
significant signals for CP violation
in the $B$-sector \cite{BaBar}, \cite{Belle}: \be \sin 2\beta =
0.75 \pm 0.09_{\rm{stat}} \pm 0.04_{\rm{syst}}~ (\rm{BaBar}),~~~
\sin 2\beta = 0.99 \pm 0.14_{\rm {stat}} \pm 0.06_{\rm{syst}}
~(\rm{Belle}).\label{sinbeta}\ee The discovery of CP violation in
the $B$ system, as reported by the BaBar and Belle Collaborations,
is a triumph for the Standard Model. There is now compelling
evidence that the phase of the CKM matrix correctly explains the
pattern of CP-violating effects in mixing and weak decays of
Kaons, charm and beauty hadrons. Specifically, the CKM mechanism
explains why CP violation is a small effect in $K$--$\bar K$
mixing ($\epsilon_K$) and $K\to\pi\pi$ decays
($\epsilon'/\epsilon$), why CP-violating effects in tree level $D$
decays are below the sensitivity of present experiments, and why
CP violation is small in $B$--$\bar B$ mixing ($\epsilon_B$) but
large in the interference of mixing and decay in $B\to J/\psi\,K$
($\sin2\beta_{J/\psi K}$).

This paper provides a review of the selected topics of the $B$
meson  decay phenomenology. Section 2 includes a brief
recapitulation of information on weak quark transitions as
described by the CKM matrix. In Section 3 we discuss $B^0-\bar
B^0$ mixing, various types of the CP-violation, specifically
description of CP asymmetries in $B$ decays to CP eigenstates, and
$\sin2\beta$ measurements.  In section 4, the determination of the
$V_{cb}$ matrix element from exclusive and inclusive semileptonic
decays of the $B$-meson is reviewed. Some phenomenological
applications are considered in Section 5. We do not attempt to
give complete references to all related literatures. By now there
are excellent lectures and minireviews that cover the subjects in
great deals \cite{Buras03}, \cite{Q01}-\cite{AB02}. We refer to
these for more details and for more complete references to the
original literature relevant to Sections 2 and 3.

\section{CP Violation in the $B$ meson decays}

{ The SM provides us with a parameterization of CP violation but
does not explain its origin. In fact, CP violation may occur in
three sectors of the SM: {\it i}) in the quark sector via the
phase of the CKM matrix, {\it ii}) in the lepton sector via the
phases of the neutrino mixing matrix, and {\it iii}) in the strong
interactions via the parameter $\theta_{\rm QCD}$.

The non observation of CP violation in the strong interactions is
a mystery (the ``strong CP puzzle''), whose explanation requires
physics beyond the SM (such as a Peccei--Quinn symmetry, axions,
etc.). Recently the possibility of CP violation in the neutrino
sector has been explored experimentally\footnote{There is by now
convincing evidence, from the experimental study of atmospheric
and solar neutrinos,
for the existence of at least two distinct frequencies of neutrino
oscillations. The evidence so far shows the mixing of
$\nu_e\to\nu_\mu$ (solar) and $\nu_{\mu}\to\nu_{\tau}$
(atmospheric) with very small mass differences and large mixing
angles. This in turn implies non-vanishing neutrino masses and a
mixing matrix, in analogy with the quark sector and the CKM
matrix.} which is the subject of many reviews (see {\it e.g.}
\cite{Altarelli} and references therein). CP violation in the
quark sector has been studied in some detail and is the subject of
this Section.
\subsection{CKM matrix}
The interactions between the quarks and gauge bosons in the SM are
illustrated in Fig.~\ref{fig:vertices}, where the vertices (a),
(b,~c) and (d) refer to weak, electromagnetic and strong
interactions, respectively.
The vertex for the charged current interaction, in which quark
flavor $i$ changes to $j$, is depicted in Fig~1(a) and has the
Feynman rule
\begin{equation}
\label{vertex} i\frac{g_2}{2\sqrt 2} V_{ij} \gamma_\mu(1-\gamma_5)
,\end{equation} where $g_2$ is the coupling constant of the
$SU(2)_L$ gauge group and $V_{ij}$ is the $ij$ element of the CKM
matrix. Eq.~(\ref{vertex}) illustrates the $V{-}A$ structure of
the charged-current interactions.

Assuming the SM with 3 generations, the network of transition
amplitudes between the charge $-1/3$ quarks $d,s,b$ and the charge
$2/3$ quarks $u,c,t$ is described by a unitary 3$\times $3 matrix
$V_{\mathrm{CKM}}$ (the CKM matrix) whose effects can be seen as a
mixing between the  $d,s,b$ quarks:
\begin{equation} \label{CKM} \left(\matrix{d ^{\,\prime}   \cr
                s ^{\,\prime} \cr
                b ^{\,\prime} \cr
            }\right)
=
    \left(\matrix{
        V_{ud}&     V_{us}&     V_{ub}\cr
        V_{cd}&     V_{cs}&     V_{cb}\cr
        V_{td}&     V_{ts}&     V_{tb}\cr
            }\right)
    \left(\matrix{
        d \cr
                s \cr
                b \cr
            }\right).
\end{equation}
\vspace{2mm} \noindent The general parameterization of $V_{CKM}$
in terms of four parameters $\theta_{ij}$ ($ij=12,\, 13,\, 23$)
and $\delta_{13}$, recommended by the Particle Data
Group~\cite{pdg}, is

\begin{equation}\left(
\begin{array}{ccc}
c_{12}c_{13} & s_{12}c_{13} & s_{13}e^{-i\delta_{13}}\\
-s_{12}c_{23}-c_{12}s_{23}s_{13}e^{i\delta_{13}} &
c_{12}c_{23}-s_{12}s_{23}s_{13}e^{i\delta_{13}} & s_{23}c_{13}\\
s_{12}s_{23}-c_{12}c_{23}s_{13}e^{i\delta_{13}} &
-c_{12}s_{23}-s_{12}c_{23}s_{13}e^{i\delta_{13}} & c_{23}c_{13}
\end{array}\right),
\label{ckmpdg}
\end{equation}
\vspace{2mm}

\noindent where  $c_{ij}=\cos \theta_{ij} $ and $s_{ij}=\sin
\theta_{ij}$. $\delta_{13}$ is the CP-violating phase parameter.

With only two families, {\it e.g.}, in a world without beauty (or
$t$ quarks), $V_{CKM}$ can always be reduced to a real form. In
case of three families one can introduce the
phase-convention-invariant form
\begin{equation}
{\rm Im}~
V_{ij}V_{kl}V_{il}^*V_{kj}^*=J\sum\limits_{m,n=1}^3\varepsilon_{ikm}\varepsilon_{jln},
\end{equation}
where $J$ is the Jarlskog invariant \cite{Jarlskog}:
\begin{equation}
\label{jarlskog}
J=c_{12}c_{23}c_{13}^2s_{12}s_{23}s_{13}\sin\delta_{13}.
\end{equation}
CP violation is proportional to $J$ and is not zero if
$\delta_{13}\neq 0$.

\begin{figure}
\begin{center}
\begin{picture}(360,90)(0,15)
\ArrowLine(0,100)(30,100)\ArrowLine(30,100)(60,100)
\Text(15,107)[b]{$i$}\Text(45,105)[b]{$j$}
\ArrowLine(100,100)(130,100)\ArrowLine(130,100)(160,100)
\Text(115,107)[b]{$i$}\Text(145,107)[b]{$i$}
\ArrowLine(200,100)(230,100)\ArrowLine(230,100)(260,100)
\Text(215,107)[b]{$i$}\Text(245,107)[b]{$i$}
\ArrowLine(300,100)(330,100)\ArrowLine(330,100)(360,100)
\Text(315,107)[b]{$i$}\Text(345,107)[b]{$i$}
\Gluon(30,100)(30,60){2.5}{5}\Text(32,55)[t]{$W^\pm$}
\Text(30,30)[t]{(a)}
\Gluon(130,100)(130,60){2.5}{5}\Text(132,55)[t]{$Z^0$}
\Text(130,30)[t]{(b)}
\Gluon(230,100)(230,60){2.5}{5}\Text(230,53)[t]{$\gamma$}
\Text(230,30)[t]{(c)}
\Photon(330,100)(330,60){2}{7}\Text(330,55)[t]{${\mathcal G}$}
\Text(330,30)[t]{(d)} \end{picture} \caption{Quark interactions
with gauge bosons. The indices $i$ and $j$ correspond to the
different flavors
 ($i=u,c,t~j=d,s,b$).} \label{fig:vertices}\end{center}
\end{figure}

\subsection{Current experimental knowledge of the CKM matrix}

Before continuing we briefly review our current experimental
knowledge of each of the CKM magnitudes. The weak mixing
parameters $V_{ud}, V_{us}$ and $V_{cs}$ are the best known
entries of the CKM matrix, but their improvement would be very
valuable  as it can lead to a better check of the unitarity of the
CKM matrix. We first consider the submatrix describing mixing
among the first two generations. The results are collected in
Table \ref{table:2_generations}.

\begin{table}
\begin{center}
\begin{tabular}{|c|c|c|}

 \hline $|V_{ij}|$   & Method & Ref. \cite{pdg}

\\ \hline
 $|V_{ud}|$     &  nuclear $\beta$ decay & $0.9734\pm0.0008$ \\
 $|V_{us}|$   &  $K\to\pi\ell\bar\nu_\ell$ &  $0.2196\pm 0.0023$ \\
 $|V_{cs}|$  & $D\to K\ell^+\nu_\ell$   & $0.996\pm0.013$ \\
   $|V_{cs}|$ & $W^+\to c\bar s$&$1.00\pm0.13$\\
   $ |V_{cd}|$ & $\nu_\mu+d\to c+\mu^-$& $0.224\pm0.016$\\

\hline
\end{tabular}
\end{center}
\caption{\label{table:2_generations} The $V_{CKM}$ submatrix
describing mixing among the first two generations. The
experimental results are taken from the Particle Data Book
\cite{pdg}.}
\end{table}

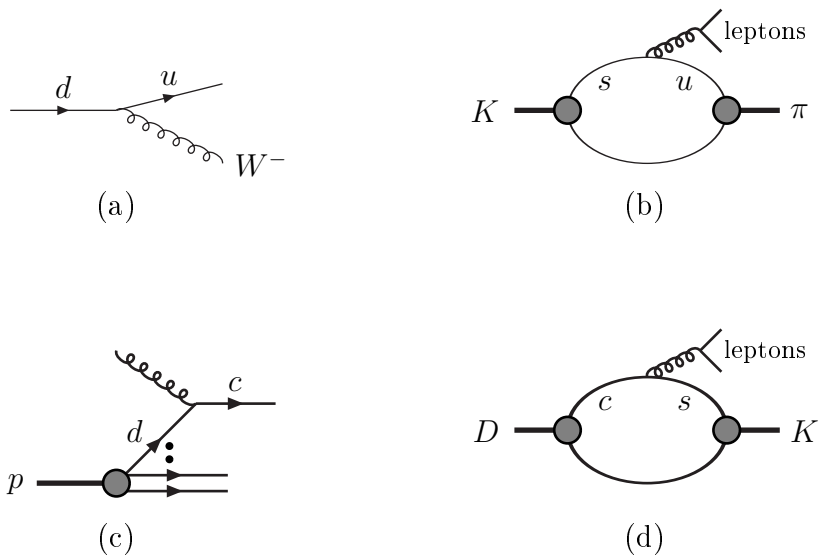
\begin{figure}[th]
\begin{center}
\begin{picture}(320,120)(0,20)
\ArrowLine(10,100)(50,100)\ArrowLine(50,100)(90,110)
\Gluon(50,100)(90,80){2}{6}
\Text(30,105)[b]{$d$}\Text(70,110)[b]{$u$}
\Text(95,80)[l]{$W^-$}\Text(50,70)[t]{(a)}
\Oval(250,100)(20,30)(0)
\SetWidth{2}\Line(220,100)(200,100)\Line(280,100)(300,100)
\SetWidth{1} \GCirc(220,100){5}{0.5}\GCirc(280,100){5}{0.5}
\Gluon(250,120)(270,130){2}{4}
\Line(270,130)(278,138)\Line(270,130)(278,122)
\Text(280,130)[l]{{\footnotesize leptons}}
\Text(195,100)[r]{$K$}\Text(305,100)[l]{$\pi$}
\Text(235,108)[b]{$s$}\Text(265,108)[b]{$u$} \Text(250,70)[t]{(b)}
\end{picture}
\begin{picture}(320,90)(0,50)
\SetWidth{2}\Line(20,80)(50,80)\SetWidth{1} \Text(15,80)[r]{$p$}
\ArrowLine(50,80)(80,110)\Text(60,100)[r]{$d$}
\ArrowLine(80,110)(110,110)\Text(95,115)[b]{$c$}
\ArrowLine(52,77)(92,77)\ArrowLine(52,83)(92,83)
\Gluon(80,110)(50,130){2}{5}
\GCirc(50,80){5}{0.5}\GCirc(70,89){1}{0}\GCirc(70,94){1}{0}
\Text(50,65)[t]{(c)}
\Oval(250,100)(20,30)(0)
\SetWidth{2}\Line(220,100)(200,100)\Line(280,100)(300,100)
\SetWidth{1} \GCirc(220,100){5}{0.5}\GCirc(280,100){5}{0.5}
\Gluon(250,120)(270,130){2}{4}
\Line(270,130)(278,138)\Line(270,130)(278,122)
\Text(280,130)[l]{{\footnotesize leptons}}
\Text(195,100)[r]{$D$}\Text(305,100)[l]{$K$}
\Text(235,108)[b]{$c$}\Text(265,108)[b]{$s$} \Text(250,65)[t]{(d)}
\end{picture}
\caption{Subprocesses from which the $V_{ud}$, $V_{us}$, $V_{cd}$,
and $V_{cs}$ elements of the CKM matrix are determined}
\label{fig:subprocesses}
\end{center}
\end{figure}

The parameter $|V_{ud}|$ is measured by studying the rates for
nuclear super-allowed and neutron $\beta$ decays. The
corresponding quark diagram is shown in Fig.
\ref{fig:subprocesses}a. Here the isospin symmetry of the strong
interactions is used to control the nonperturbative dynamics,
since the operator $\bar d\gamma^\mu(1-\gamma^5)u$ is a partially
conserved current associated with a generator of chiral
$SU(2)_L\times SU(2)_R$. The present  data yield the value of
$|V_{ud}|$ with accuracy of $0.1\%$.

The parameter $|V_{us}|$ is essentially derived from
$K\to\pi\ell\bar\nu_\ell$ decay (see Fig. \ref{fig:subprocesses}b)
while the hyperon $\Lambda\to p\ell\bar\nu_\ell$ decay plays an
ancillary role. Here chiral $SU(3)L\times SU(3)_R$ symmetry must
be used in the hadronic matrix elements, since a strange quark is
involved. Because the $m_s$ corrections are larger, $|V_{us}|$ is
only known to $1\%$.

The CKM elements involving the charm quark are not so well
measured. One can extract $|V_{cd}|$ from deep inelastic neutrino
scattering on nucleons, using the process $\nu_\mu+d\to c+\mu^-$
(see Fig. \ref{fig:subprocesses}c). This inclusive process may be
computed perturbatively in QCD, leading to a result with accuracy
at the level of~$10\%$. One way to extract $|V_{cs}|$ is to study
the decay $D\to K\ell^+\nu_\ell$ (see Fig.
\ref{fig:subprocesses}d). In this case there is no symmetry by
which one can control the matrix element $\langle K|\bar
s\gamma^\mu(1-\gamma^5)c|D\rangle$, since flavor $SU(4)$ is badly
broken.  One is forced to resort to models for these matrix
elements.  The error estimate in reported value should probably be
taken to be substantially larger.
An alternative is to measure $V_{cs}$ from inclusive processes at
higher energies.  For example, one can study the branching
fraction for $W^+\to c\bar s$, which can be computed using
perturbative QCD.  The result quoted in Table
\ref{table:2_generations} is
consistent with the model-dependent measurement. In this case the
error
is largely experimental, and is unpolluted by hadronic physics.


The elements of $V_{ij}$ involving the third generation are, for
the most part, harder to measure accurately.  The branching ratio
for $t\to b\ell^+\nu$ can be analyzed perturbatively, but the
experimental data are not very good.  Measurements of the
$b$-fraction in top quark decays by CDF and D0 result in the
rather loose restriction on $|V_{tb}|$
\begin{equation}
\label{vtb}
  |V_{tb}|=
  (0.99\pm 0.15)\times (|V_{td}|^2+|V_{ts}|^2+|V_{tb}|^2).
\end{equation}

There are as yet no direct extractions of $|V_{td}|$ or
$|V_{ts}|$. One can use the experimental data  for the ratio
${\cal B}(B \to X_s \gamma)/{\cal B}(B\to X_c\ell\nu_{\ell})$ and
the theoretical prediction for ${\cal B}(B \rightarrow X_s
\gamma)$  in order to directly determine the combination $|V_{tb}
V_{ts}^*|/|V_{cb}|$. In this way averaging the CLEO \cite{CLEObsg}
and ALEPH data \cite{ALEPHbsg}, one obtains (for details see
\cite{greubhurthQCD})
\begin{equation}
 \frac{|V_{ts}^* V_{tb}|}{|V_{cb}|} =0.93 \pm 0.10, \end{equation}
 where all the errors were added in
quadrature. Using $|V_{tb}|$ from (\ref{vtb}) and $|V_{cb}|=(40.6
\pm 1.1)\times 10^{-3}$ extracted from semileptonic $B$ decays
(see Section 3), one obtains
\begin{equation} \label{vts} |V_{ts}|=0.038 \pm 0.007.
\end{equation} This
is probably the most direct determination of this CKM matrix
element. With an improved measurement of ${\cal B}(B \to X_s
\gamma)$ and $V_{tb}$, one expects to reduce the present error on
$|V_{ts}|$ by a factor of 2 or even more.

This leaves us with the matrix elements $V_{ub}$ and $V_{cb}$, for
which we need an understanding of $B$ meson decay. This issue will
be discussed in Section 3.

\subsection{The Wolfenstein parameterization}

{\noindent The parametrization ({\ref{ckmpdg}) is general, but
awkward to use. For most practical purposes it is sufficient to
use a simpler, but approximate Wolfenstein parametterization
\cite{wolf-prl}, which, following the observed hierarchy between
the CKM matrix elements, expands the CKM matrix in terms of the
four parameters
\begin{equation}\label{eq:newparametrization}
s_{12}=\lambda,~~~s_{23}=A\lambda^2,~~~s_{13}e^{-i\delta}=A\lambda^3(\rho-i\eta),
\end{equation}
\noindent with $\lambda$ being the expansion parameter. In terms
of these parameters one finds with accuracy up to ${\cal
O}(\lambda^6)$ \cite{BLO94}
\begin{equation} V_{ud}=1-\frac{1}{2}\lambda^2-\frac{1}{8}\lambda^4,~~~V_{us}=\lambda+{\cal O}(\lambda^7),
\end{equation}
\begin{equation} V_{cd}=-\lambda+\frac{1}{2}
A^2\lambda^5 [1-2 (\rho+i \eta)],~~V_{cs}=
1-\frac{1}{2}\lambda^2-\frac{1}{8}\lambda^4(1+4 A^2),
\end{equation}
\begin{equation}\label{VUS} ~~~ V_{cb}=A\lambda^2+{\cal O}(\lambda^8),~~V_{ub}=A
\lambda^3 (\rho-i \eta),~~~
\end{equation}
\begin{equation}\label{barred_quantities}
 V_{ts}= -A\lambda^2+\frac{1}{2}A\lambda^4[1-2 (\rho+i\eta)]
,~~ V_{td}=A\lambda^3(1-\bar\rho-i\bar\eta),~V_{tb}=1-\frac{1}{2}
A^2\lambda^4, \qquad
\end{equation}
The barred quantities in  (\ref{barred_quantities}) are
\begin{equation}
\bar\rho=\rho(1-\frac{\lambda^2}{2}),~~~\bar\eta=\eta(1-\frac{\lambda^2}{2}).
\end{equation}
In the Wolfenstein parameterization, the CKM matrix is written
with accuracy up to ${\cal O}(\lambda^4)$
\begin{equation}
V_{\mathrm{CKM}}=\left(
\begin{array}{ccc}
1-\frac{\lambda^2}{2} & \lambda & A\lambda^3 (\rho - i \eta)\\
-\lambda & 1-\frac{\lambda^2}{2}\rule{0pt}{14pt} & A\lambda^2\\
A\lambda^3 (1- \rho - i \eta) & - A\lambda^2 \rule{0pt}{14pt}& 1
\end{array}\right).
\label{eq:wolfenstein}\end{equation} This parameterization
corresponds to a particular choice of phase convention which
eliminates as many phases as possible and puts the one remaining
complex phase in the matrix elements $V_{ub}$ and $V_{td}$.

The parameters $\lambda$  is known with good precision:
\begin{equation}
     \lambda=\sin\theta_{12}=0.2237 \pm 0.0033.
    \label{eq:lambda}
\end{equation}
The rate of the allowed $b\to c$ decay leads to a determination of
the combination $A\lambda^2$ :
\begin{equation}
    A\lambda^{2}=V_{cb}= (41.0 \pm 1.6)\times 10^{-3} \, .
    \label{eq:Al2}
\end{equation}
The problem of determining $\rho$ and $\eta$ is best seen in the
light of the unitarity relation.
\begin{figure}
\centerline{\includegraphics[height=90mm,keepaspectratio=true]{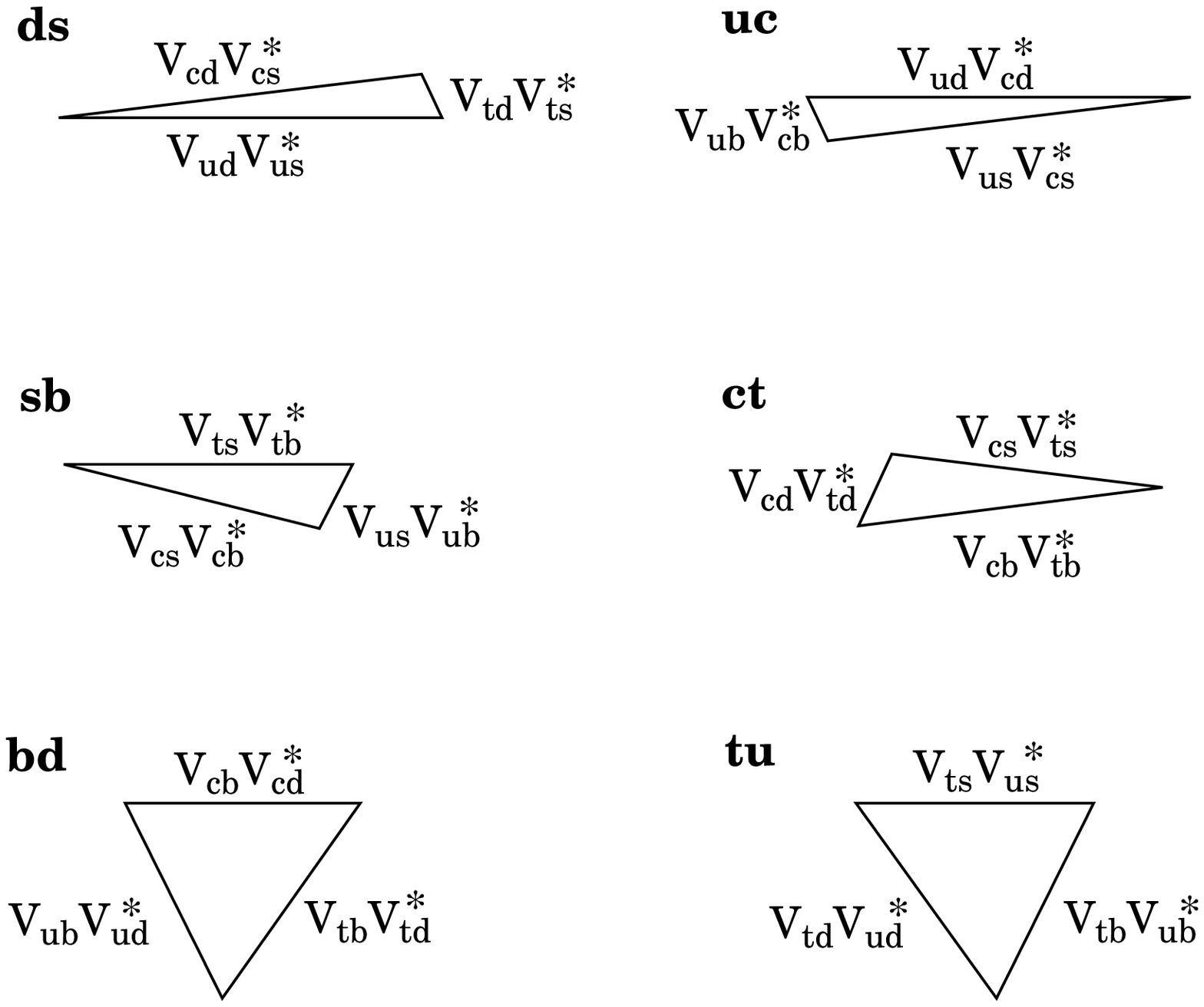}}
\caption{\label{6_triangles}Six unitarity triangles}
\end{figure}
\subsection{The unitarity triangle}

The unitarity of the CKM matrix implies various relations between
its elements:
\begin{equation}
\label{ur} \sum\limits_k V_{ij}V_{kj}^*=\delta_{ij}
\end{equation}
The {\it
 unitarity triangles} are geometrical representations in the complex plane of the
 six equations (\ref{ur}) with $i\ne k$. It is a trivial fact that any relationship
 of the form of a sum of three complex numbers equal to zero can be drawn as a closed triangle,
 see fig. \ref{6_triangles}.

 All the unitarity triangles have the same area, $J/2$. However,
while the triangles have the same area, they are of very different
shapes: {\it e.g.} {\bf ds} triangle has two sides of order
$\lambda$ and one of order $\lambda^4$, while {\bf sb} triangle
has larger sides of order $\lambda^2$ and the small side of order
$\lambda^5$ giving an angle of order $\lambda^2$. It would be very
difficult to measure the area using such triangles. This leaves us
with the {\bf bd} triangle corresponding to the
 relation
\begin{equation}\label{2.87h}
V_{ud}^{*}V_{ub} + V_{cd}^{*}V_{cb} + V_{td}^{*}V_{tb} =0,
\end{equation}
in which all sides are of order $\lambda^3$. The relation
(\ref{2.87h}) is phenomenologically especially interesting as it
involves simultaneously the elements $V_{ub}$, $V_{cb}$, and
$V_{td}$, which are under extensive discussion at present. To an
excellent accuracy $V_{cd}V_{cb}^*$ is
\begin{equation}| V_{cd}V_{cb}^*|=A\lambda^3+{\cal
O}(\lambda^7)\end{equation} Rescale all terms in (\ref{2.87h}) by
$A \lambda^3$ and put the vector $V_{cd}^{}V_{cb}^*$ on the real
axis. The coordinates of the remaining vertex correspond to the
$\rho$ and $\eta$ parameters or, in an improved version
\cite{BLO94}, to $\bar\rho=(1-\lambda^2/2)\rho$ and
$\bar\eta=(1-\lambda^2/2)\eta$. The corresponding triangle is
shown in Fig. \ref{ckm_tri}.

\noindent  The angles $\alpha$, $\beta$, and $\gamma$ (according
to the BaBar collaboration, also known as, respectively, $\phi_2$,
$\phi_1$, and $\phi_3$ according to the Belle collaboration) are
defined as follows:
\begin{equation}
\alpha=arg\left(-\frac{V_{td}V_{tb}^*}{V_{ud}V_{ub}^*}\right),~~
\beta=arg\left(-\frac{V_{cd}V_{cb}^*}{V_{td}V_{tb}^*}\right),~~
\gamma=arg\left(-\frac{V_{ud}V_{ub}^*}{V_{cd}V_{cb}^*}\right).
\end{equation}
The angles $\beta$ and $\gamma=\delta_{\rm CKM}$ of the unitarity
triangle are related directly to the complex phases of the
CKM-elements $V_{td}$ and $V_{ub}$, respectively, through
\be\label{e417} V_{td}=|V_{td}|e^{-i\beta},\quad
V_{ub}=|V_{ub}|e^{-i\gamma}. \ee
 The lengths $R_u$ and $R_t$ are
\begin{equation} R_u=|\rho+i\eta|=\left|
\frac{V_{ud}V_{ub}^*}{V_{cd}V_{cb}^*}\right|,~~R_t=|1-\rho-i\eta|=\left|
\frac{V_{td}V_{tb}^*}{V_{ud}V_{ub}^*}\right|.\end{equation}Since
the area of the unitarity triangle is $\eta/2$, a non-flat
triangle implies $CP$ violation.

Within the SM the measurements of four CP {\it conserving } decays
sensitive to $|V_{us}|$, $|V_{ub}|$, $|V_{cb}|$ and $|V_{td}|$ can
tell us whether CP violation ($\bar\eta \not= 0$) is predicted in
the SM. This fact is often used to determine the angles of the
unitarity triangle by measurements of $CP$ conserving quantities.
The length of one side, $|\rho+i\eta|$, is extracted from a
determination of $V_{ub}$, e.g from the rates of the forbidden
$b\rightarrow u$ semileptonic transitions.  CP-violating
$K^0$--$\bar K^0$ mixing is dominated by $\bar s d \to \bar d s$
with virtual $t \bar t$ and $W^+ W^-$ intermediate states.  It
constrains Im $(V_{td}^2) \sim \bar \eta (1 - \bar \rho)$, giving
a hyperbolic band in the $(\bar \rho, \bar \eta)$ plane. The
important constraint comes from $B^0-\bar B{}^0$ mixing, which is
mediated by the  operator $\bar b\gamma^\mu(1-\gamma^5)d\, \bar
b\gamma^\mu(1-\gamma^5)d$. In the SM, this operator is generated
by the loop diagram for $\bar b d \to \bar d b$ (see Fig.
\ref{boxes}), with a coefficient proportional to
$|V_{td}{}^*V_{tb}|^2$.  The phenomenological parameter $\Delta
m_d$ is precisely measured, see subsection \ref{timeevolution}.
However, as in the case of $K^0-\bar K{}^0$ mixing, relating this
number to fundamental quantities requires hadronic matrix elements
which are difficult to compute.

The above mentioned determinations (schematically shown in Fig.
\ref{ckm_tri}) point to a non-flat triangle, i.e. to the presence
of a certain amount of $CP$ violation.

\begin{figure}
\centerline{\includegraphics[height=70mm,keepaspectratio=true]{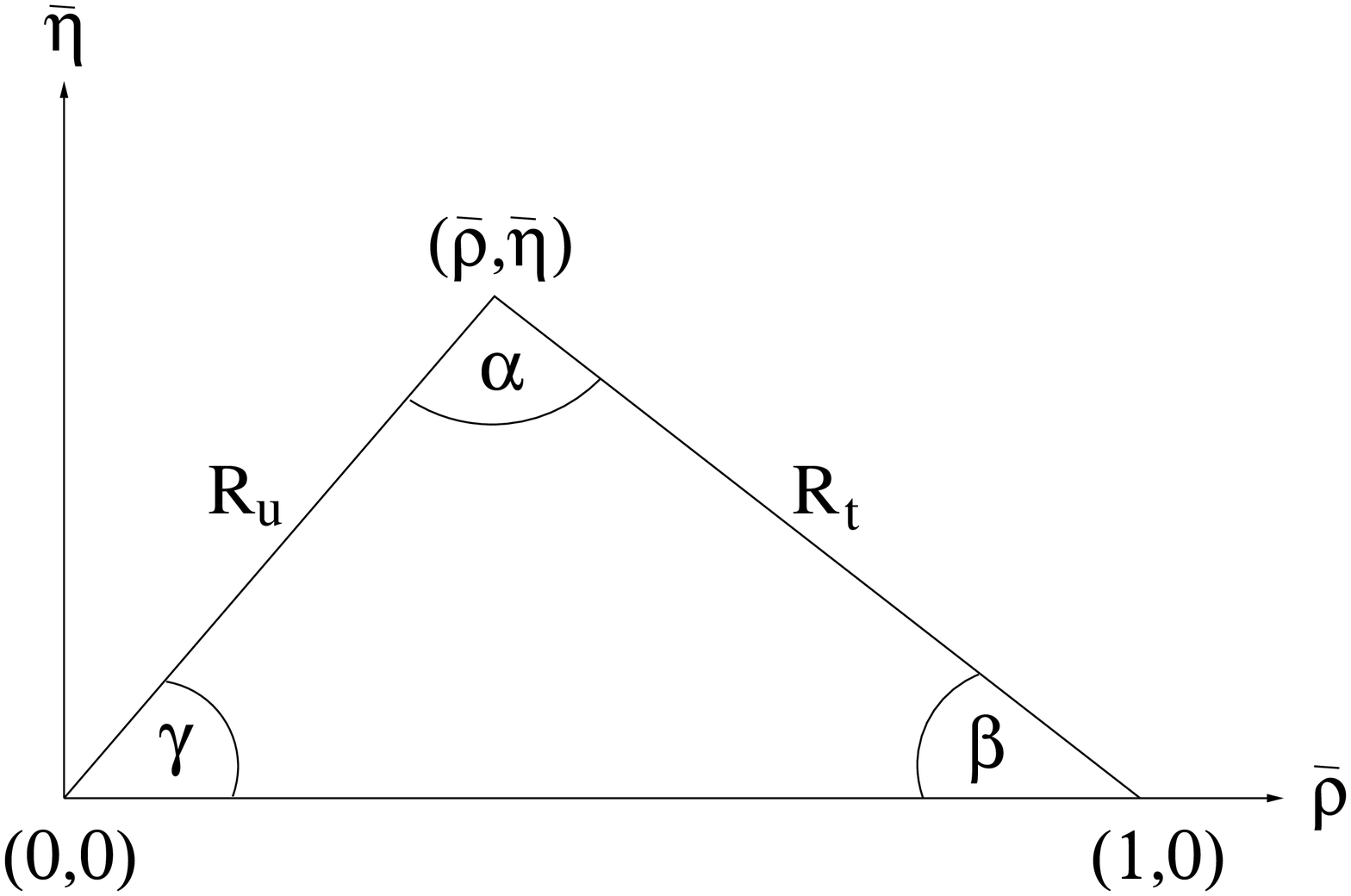}}
\caption{\label{ckm_tri}Unitarity triangle corresponding to Eq.
(\ref{2.87h})}
\end{figure}
\begin{figure}
\centerline{\includegraphics[height=40mm,keepaspectratio=true]{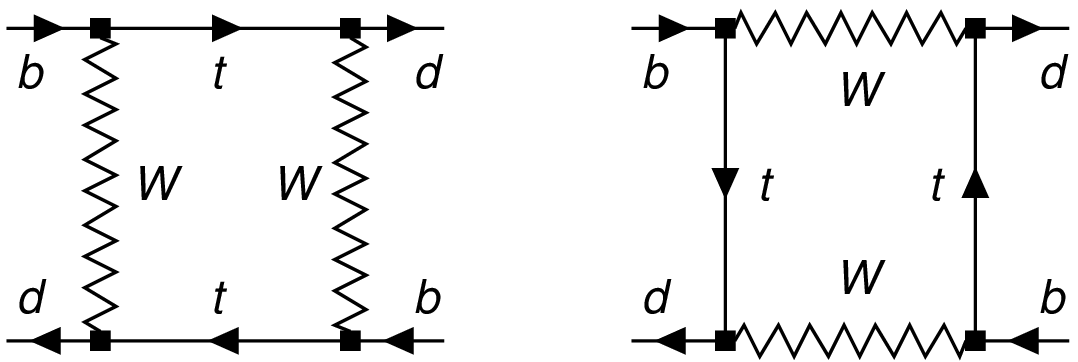}}
\caption{\label{boxes}SM box diagrams including $B^0-{\bar B}^0$
mixing.}
\end{figure}

\begin{figure}
\hspace*{20mm}\includegraphics[height=3in]{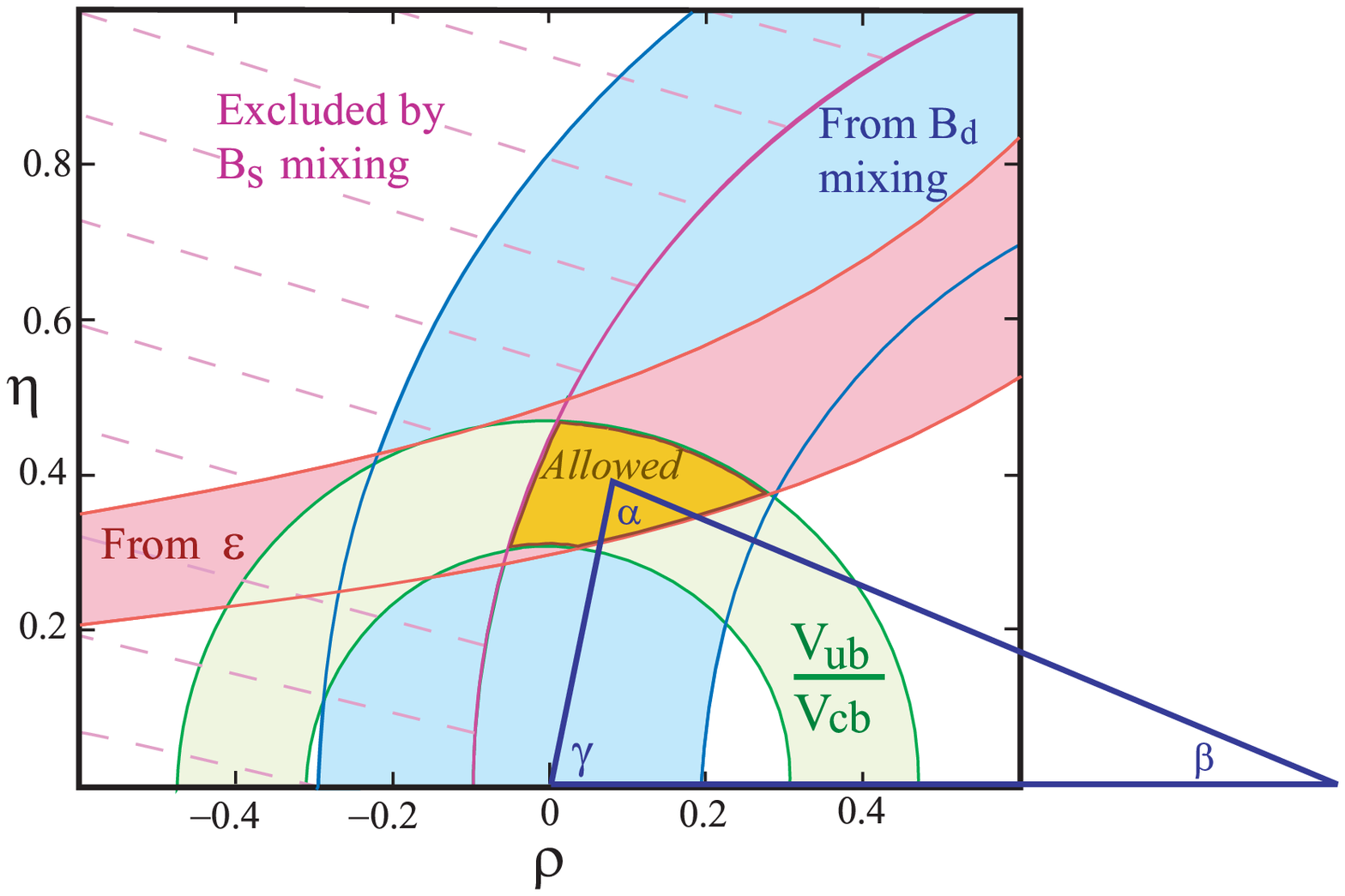}
\caption{\label{ckm_tri}Schematic  determination of the unitarity
triangle.}
\end{figure}


Several  global analyses~of the unitarity triangle have been
performed, combining measurements of $|V_{cb}|$ and $|V_{ub}|$ in
semileptonic $B$ decays, $|V_{td}|$ in $B$--$\bar B$ mixing, and
the CP-violating phase of $V_{td}^2$ in $K$--$\bar K$ mixing and
$B\to J/\psi\,K$ decays, see {\it e.g.} Refs.
\cite{Hocker:2001xe}, \cite{V03}. The values obtained at 95\%
confidence level are
\begin{equation}
\bar\rho=0.178\pm 0.046,~~ \bar\eta=0.341\pm 0.028.\end{equation}
The corresponding results for the angles of the unitarity triangle
are
\begin{equation}
\sin 2\beta=0.705^{+0.042}_{-0.032},~~~\sin2\alpha=-0.19\pm
0.25,~~~ \gamma=(61.5\pm 7.0)^\circ.\end{equation} These studies
have established the existence of a CP-violating phase in the top
sector of the CKM matrix, i.e., the fact that
$\mbox{Im}(V_{td}^2)\propto\bar\eta\ne 0$.

\section{Express review of the phenomenology of CP violation}

\subsection{Time evolution and mixing}
\label{timeevolution}
We first list the necessary formulae to
describe $B^0-{\bar B}^0$ mixing. The formulae are general and
apply to both $B_d^0$ and $B_s^0$ mesons  although with different
values of parameters. In the following, we use the standard
convention that $B^0$ (${\bar B}^0$) contains $\bar b$ antidiquark
($b$ quark).

Once $CP$ is not a symmetry of the theory one must allow a more
general form for the two mass eigenstates of neutral but flavored
mesons. These two states are usually defined as $B_H$ and $B_L$
where the $H$ and $L$ stand for heavier and less heavy mesons. The
light and heavy mass eigenstates can be written as linear
combinations of $B^0$ and $\bar B^0$: \bea\label{defqp}
|B_H\rangle&=&p|{B^0}\rangle+q|{\bar B^0}\rangle,
\nonumber\\|B_L\rangle&=&p|{B^0}\rangle-q|{\bar B^0}\rangle, \eea
with \be\label{norqp} |q|^2+|p|^2=1. \ee The phase convention used
here is $CP~|B^0> = |\bar B^0>$ that makes the phase of $q$ a
meaningful quantity. The mass difference $\Delta m_B$ and the
width difference $\Delta\Gamma_B$ are defined as follows:
\be\label{DelmG} \Delta m_B\equiv M_H-M_L,\ \ \
\Delta\Gamma_B\equiv\Gamma_H-\Gamma_L. \ee The average mass and
width are given by \be\label{aveMG} m_B\equiv{M_H+M_L\over2},\ \ \
\Gamma_B\equiv{\Gamma_H+\Gamma_L\over2}. \ee

The time evolution of the mass eigenstates is simple:
\bea\label{temes} |B_H(t)\rangle&=&
e^{-iM_Ht}e^{-\Gamma_Ht/2}|B_H(0)\rangle,\nonumber\\
|B_L(t)\rangle&=& e^{-iM_Lt}e^{-\Gamma_Lt/2}|B_L(0)\rangle. \eea
In the presence of flavor mixing the time evolution of $B^0$ and
$\bar B^0$ is more complicated. An initially produced $B^0$ or
${\bar B}^0$ evolves in time into a superposition of $B^0$ and
${\bar B}^0$. Let $B^0(t)$ denotes the state vector of a $B$ meson
which is tagged as $B^0$ at time $t=0$, {\it i.e.}
$|B^0(t=0)>=B^0$. Likewise ${\bar B}^0(t)$ represent a $B$ meson
initially tagged as ${\bar B}^0$. The time evolution of these
states is governed by a Schr\"odinger-like equation \vspace{1mm}

\be\label{Schro} i{d\over dt}\pmatrix{B^0(t)\cr \bar B^0(t)\cr}=
\left(M-{i\over2}\Gamma\right)\pmatrix{B^0(t)\cr \bar B^0(t)\cr},
\ee \vspace{1mm}

\noindent where the mass matrix  $M$ and decay matrix $\Gamma$ are
time independent, hermitian $2\times 2$ matrices written in the
basis of the two flavor eigenstates. Both M and $\Gamma$ are
complex with $M_{21}=M_{12}^*$, $\Gamma_{21}=\Gamma_{12}^*$
(hermicity), $M_{11}=M_{22}^*$, $\Gamma_{11}=\Gamma_{22}^*$ (CPT).
$M_{12}$ is the dispersive part of the transition amplitude from
$B^0$ to $\bar B^0$, while $\Gamma_{12}$ is the absorptive part of
that amplitude. The  off-diagonal terms in (\ref{Schro})  are
induced by  $|\Delta B|=2$ transitions, so that the mass
eigenstates of the neutral $B$ mesons that are defined as
eigenvectors of $M-i\Gamma/2$ are different from the flavor
eigenstates $B^0$ and ${\bar B}^0$. Solving the eigenvalue
equation \vspace{1mm}

\be\label{eigenvalue} \left(M-{i\over2}\Gamma\right)\pmatrix{p\cr
 q\cr}=\lambda\pmatrix{p\cr q\cr}, \ee \vspace{1mm}

\noindent we obtain~
\be
\lambda_{\pm}=M_{11}-\frac{i}{2}\Gamma_{11}\pm\sqrt{(M_{12}-\frac{i}{2}
\Gamma_{12})(M_{12}^*-\frac{i}{2}\Gamma_{12}^*)},
\label{lambda}\ee \vspace{1mm}

\noindent where $\lambda_+=\lambda_H$ and $\lambda_-=\lambda_L$.
The off-diagonal (or mixing) elements are calculated from Feynman
Diagrams that can convert one flavor eigenstate to the other.  In
the Standard Model these are dominated by the one loop box
diagrams, shown in Fig. \ref{boxes}.

To find the time evolution of $B^0(t)$ and ${\bar B}^0(t)$ we
invert Eqs. (\ref{defqp}) to express $B^0$ and ${\bar B}^0$ in
terms of the mass eigenstates $B_L$ and $B_H$ and use their time
evolution, Eqs. (\ref{temes}).  Then the evolution of the
eigenstates (\ref{defqp}) of well-defined masses  $M_{\pm} = {\rm
Re}(\lambda_{\pm})$ and decay widths $\Gamma_{\pm} = -2{\rm
Im}(\lambda_{\pm})$ is given by the phases
$\exp(-i\lambda_{\pm}t)$ where  \be\label{phase} \lambda_{\pm}=
M_{\pm}-i\frac{1}{2}\Gamma_{\pm}t.\ee \noindent Using Eq.
(\ref{lambda}) one obtains \bea\label{eveq} (\Delta
m_B)^2-{1\over4}(\Delta\Gamma_B)^2=(4|M_{12}|^2-|\Gamma_{12}|^2),~~~
\Delta m_B\Delta\Gamma_B=4{\rm Re}(M_{12}\Gamma_{12}^*),\eea and
 \be\label{solveqp} {q\over
p}=\sqrt{\frac{M_{12}^*-\frac{i}{2}\Gamma_{12}^*}{M_{12}-\frac{i}{2}\Gamma_{12}}}
. \ee \vspace{1mm}The time evolution of a pure $| B{^0}\rangle$ or
$|\overline B{^0}\rangle$ state at $t=0$ is thus given by
\begin{eqnarray}
| B{^0}(t)\rangle &=& g_+(t) \,| B{^0}\rangle
                     + \frac{q}{p} \, g_-(t) \,|\overline B{^0}\rangle \,,
\label{time_evol1}
\\
|\overline B{^0}(t)\rangle &=& g_+(t) \,|\overline B{^0}\rangle
                     + \frac{p}{q} g_-(t) \,| B{^0}\rangle \,,
\label{time_evol2}
\end{eqnarray}
where
\begin{equation} 
g_{\pm}(t) = \frac{1}{2} \left(e^{-i\lambda_+ t} \pm
e^{-i\lambda_- t}  \right) \,.
\end{equation} 

The flavor states remain unchanged  or oscillate into each other
with time-dependent probabilities proportional to
\begin{equation}
\left| g_{\pm}(t)\right|^2 = \frac{e^{-\Gamma_B t}}{2} \left[
\cosh\!\left( \frac{\Delta\Gamma_B}{2}\,t\right) \pm \cos(\Delta
m\,t)\right]. \label{cosh_cos}
\end{equation}

One can expect that $\Delta\Gamma_B/\Gamma_B\ll 1$ and
$|\Gamma_{12}/M_{12}|\ll 1$ for $B^0-\bar B^0$ mixing (this is not
the case for $K^0-\bar K^0$ mixing). The reason is that, on the
one hand, it is experimentally known that $\Delta
m_B/\Gamma_B\approx 0.7$. On the other hand, the difference in
widths is produced by decay channels common to $B^0$ and $\bar
B^0$. The branching ratios for such channels are at or below the
level of $10^{-3}$. Since various channels contribute with
differing signs, one expects that their sum does not exceed the
individual level. Hence, we can safely assume that \be
\frac{\Delta\Gamma_B}{\Gamma_B}\sim\left|
\frac{\Gamma_{12}}{M_{12}}\right|={\cal O}(10^{-2}).\ee

To leading order in $|\Gamma_{12}/M_{12}|$, Eqs. (\ref{eveq}) and
(\ref{solveqp}) can be written as \be\label{eveqB} \Delta
m_B=2|M_{12}|,~~~\Delta\Gamma_B= 2{\rm
Re}\frac{M_{12}\Gamma_{12}^*}{|M_{12}|},\ee and \label{solveqpB}
\be\frac{q}{p}=-\frac{M_{12}^*}{|M_{12}|}. \ee Note that the two
mass eigenstates  do not have to be orthogonal, in fact in general
they will not be so, unless $|q/p|=1$.

\subsection{The box diagram}

The mass difference $\Delta m_{B}$ is is a measure of the
oscillation frequency to change from $B^0$ to $\bar B^0$ and vise
versa. Because the long distance contributions for $B^0-\bar B^0$
mixing are small (in contrast with the situation for $\Delta m_K$)
$\Delta m_B$ and $\Delta m_B$ are very well approximated by the
relevant box diagram. Since $m_{u,c}\ll m_t$ the only
non-negligible contributions to $M_{12}$ and $\Gamma_{12}$ are
from box diagrams involving two top quarks, the charm and mixed
top-charm contributions are entirely negligible.

The dispersive ($M_{12}$) and absorptive ($\Gamma_{12}$) parts of
top-mediated box diagrams are given by

\begin{equation}
\label{m12} M_{12} = - \frac{
           G_F^2 m_W^2 \eta_B m_{B} B_{B} f_{B}^2}{12\pi^2}
           \, S_0(\frac{m_t^2}{m_W^2}) \, (V_{td}^* V_{tb}^{})^2,
           \end{equation}
           \be\label{gamma12}
\Gamma_{12}  =  \frac{
           G_F^2 m_b^2 \eta'_B m_{B} B_{B}
           f_{B}^2}{8\pi}\,
            (V_{td}^* V_{tb}^{})^2,
\ee
\noindent where $m_W$ is the $W$ boson mass and  $m_i$ is the mass
of quark $i$.
 The factor $f_{B}$ is the vacuum-to-one-meson matrix
element of the axial current, which arises in the naive
approximation obtained by splitting the matrix element into
two-quark terms and inserting the vacuum state between them. This
is known as the vacuum-insertion approximation. The quantity
$B_{B}$ is simply the correction factor between that approximate
answer and the true answer. It can be estimated in various model
calculations. The QCD corrections $\eta_B$ and $\eta'_B$ are of
order unity ($\eta_B=0.55\pm0.01$). The known function $S_0(x_t)$
can be approximated very well with \be
S_0(x_t)=0.784\,x_t^{0.76}\ee For more details and further
references see \cite{Buras_Fleischer_HeavyFlavorsII}.

New physics usually takes place at a high energy scale and is
relevant to the short distance part only. Therefore, the SM
estimate in Eqs. (\ref{m12}) and (\ref{gamma12}) remains valid
model independently. Combining (\ref{m12}) and (\ref{gamma12}), we
obtain that \be\left|\frac{\Gamma_{12}}{M_{12}}\right|\approx
\frac{3\pi m_b^2}{2m_W^2S(x_t)}\approx 5\times 10^{-3},\ee for
$m_b=4.25$ GeV, $m_W=80$ GeV and $m_t=174$ GeV, which confirms our
previous order of magnitude estimate.
$2\phi_B=\arg(V_{td^*}V_{tb})^2$ is a CP violating phase. The
phase of $\Gamma_{12}$ is given by \be\arg\Gamma_{12}=2\phi_B,\ee
where $\phi_B$ is a CP violating phase:
\be2\phi_b=\arg(V_{td}^*V_{tb})^2=2\beta,\ee the phase of $M_{12}$
is \be\arg M_{12}=2\beta+\pi.\ee  The leading correction to this
result is proportional to $(m_c/m_b)^2$, more precisely, the phase
difference between $M_{12}$ and $\Gamma_{12}$ is \be \arg
M_{12}-\arg\Gamma_{12}=
\pi+\frac{8}{3}\left(\frac{m_c}{m_b}\right)^2\times\frac{\eta}{(1-\rho)^2+\eta^2},\ee
{\it i.e.} {\it i.e.}  $M_{12}$ and $\Gamma_{12}$ are almost
antiparallel. For the $B$ system this leads to \be\label{qpforB}
\left(\frac{q}{p}\right)_B= e^{-2i\phi_B}. \ee At leading order in
$\lambda$ and next-to-leading order in QCD, one finds \be \Delta
m_B=1.30\frac{G_F^2M_W^2}{6\pi^2}\cdot m_{B} \cdot
f_{B}^2B_{B}\cdot A^2\lambda^6\cdot[(1-\rho)^2+\eta^2]. \ee

After surprising discovery of large $B^0-{\bar B}^0$ mixing by the
Argus collaboration  many $B^0-\bar B^0$ oscillations experiments
were performed by the
different experimental groups (for the complete list of references
see \cite{S02}). Although a variety of techniques have been used,
the individual $\Delta m_B$ results obtained at high-energy
colliders have remarkably similar precision. Their average is
compatible with the recent and more precise measurements from
asymmetric $B$ factories. Before being combined, the measurements
are adjusted on the basis of a common set of input values,
including the $b$-hadron lifetimes and fractions.
Combining all published measurements  PDG 2002 quotes the value of
\be\Delta m_B = {\rm 0.489 \pm 0.005 (stat) \pm 0.007
(syst)}~\hbox{ps}^{-1}.\ee

\subsection{CP violating effects}.For the $B$ mesons it is useful to make a classification
of CP-violating effects that is more transparent than the division
into {\it indirect} and {\it direct} CP violation usually
considered for the Kaon sector. In the SM, there are several
possible ways of $CP$ violation. The first, seen for example in
$K$ decays, occurs if $|q/p| \neq 1$. It is very clear in this
case that no choice of phase conventions can make the two mass
eigenstates be $CP$ eigenstates. This is generally called
$CP$-violation in the mixing. A second possibility is $CP$
violation in the decay, or direct $CP$ violation, which requires
that two $CP$-conjugate processes to have differing absolute
values for their amplitudes. The third option is $CP$ violation in
the interference between decays with and without mixing. We shall
consider these cases step by step. A detailed presentation can be
found {\it e.g.} in Ref. \cite{nyr01}

\subsubsection{CP Violation in Mixing} This type of CP violation
results from the mass eigenstates being different from the CP
eigenstates, and requires a relative phase between $M_{12}$ and
$\Gamma_{12}$, {\it i.e.} $ |q/p|\neq1$.  CP violation in mixing
has been observed in the neutral $K$ system (${\rm Re}~
\varepsilon_K\neq0$). For the neutral $B$ system, this effect can
be best isolated by measuring the asymmetry in semileptonic
decays: \be\label{mixexa} A_{\rm SL}={\Gamma(\bar
B^0(t)\to\ell^+\nu X)- \Gamma(B^0(t)\to\ell^-\nu X)\over
\Gamma(\bar B^0(t)\to\ell^+\nu X)+ \Gamma(B^0(t)\to\ell^-\nu X)}.
\ee The final states in (\ref{mixexa}) contain ``wrong charge''
leptons and can be only reached in the presence of $B^0-{\bar
B}^0$ mixing. As the phases in the $B^0-{\bar B}^0$ and ${\bar
B}^0- B^0$ transitions differ from each other, a non-vanishing CP
asymmetry follows. Specifically, for the time-integrated CP
asymmetry one obtains \be\label{Absqp} A_{\rm SL}={\rm
Im}~\frac{\Gamma_{12}}{M_{12}}\approx-1.4\times10^{-3}{\eta\over
(1-\rho)^2+\eta^2}. \ee The suppression by a factor of ${\cal
O}(10)$ of $a_{\rm SL}$ compared to $|\Gamma_{12}/M_{12}|$ comes
from the fact that the leading contribution to $\Gamma_{12}$ has
the same phase as $M_{12}$. Consequently, $A_{\rm SL}= {\cal
O}(m_c^2/m_t^2)$. The CKM factor does not give any further
significant suppression, \be {\rm Im}\frac{V_{cb}V_{cd}^*}{
V_{tb}V_{td}^*}={\cal O}(1).\ee In contrast, for the $B_s$ system,
where the same expressions holds except that $V_{cd}/V_{td}$ is
replaced by $V_{cs}/V_{ts}$, there is an additional CKM
suppression from \be {\rm Im}\frac{V_{cb}V_{cs}^*}{
V_{tb}V_{ts}^*}={\cal O} (\lambda^2).\ee To estimate  $A_{\rm SL}$
in a given model, one needs to calculate $M_{12}$ and
$\Gamma_{12}$. This involves some hadronic uncertainties, in
particular in the hadronization models for $\Gamma_{12}$.

The asymmetry $A_{SL}$ has been searched for in several
experiments, with sensitivity at the level of $10^{-2}$ giving a
world average of \be\label{aveasl} A_{\rm
SL}=(0.2\pm1.4)\times10^{-2}. \ee

\subsubsection{CP violation in decay}  We define the decay
amplitudes $A_f$ and $\bar A_f$ according to \be\label{defAf}
A_f=\langle f|{\cal H}_d|B^0\rangle,\ \ \ \bar A_f=\langle f|{\cal
H}_d|\bar B^0\rangle, \ee where ${\cal H}_d$ is the decay
Hamiltonian.

CP relates $A_f$ and $\bar A_{\bar f}$. There are two types of
phases that may appear in $A_f$ and $\bar A_{\bar f}$. Complex
parameters in any Lagrangian term that contributes to the
amplitude will appear in complex conjugate form in the
CP-conjugate amplitude. Thus their phases appear in $A_f$ and
$\bar A_{\bar f}$ with opposite signs. In the SM, these phases
occur only in the mixing matrices
, hence these are often called ``weak phases''. The weak phase of
any single term in $A_f$ is convention dependent. However the
difference between the weak phases in two different terms in $A_f$
is convention independent because the phase rotations of the
initial and final states are the same for every term. A second
type of phase can appear in scattering or decay amplitudes even
when the Lagrangian is real. Such phases do not violate CP and
they appear in $A_f$ and $\bar A_{\bar f}$ with the same sign.
Their origin is the possible
contributions from coupled channels. Usually the dominant re
scattering is due to strong interactions and hence the designation
``strong phases'' for the phase shifts so induced. Again only the
relative strong phases of different terms in a scattering
amplitude have physical content, an overall phase rotation of the
entire amplitude has no physical consequences. Thus it is useful
to write each contribution to $A$ in three parts: its magnitude
$A_i$; its weak phase term $e^{i\phi_i}$; and its strong phase
term $e^{i\delta_i}$. Then, if several amplitudes contribute to
$B\to f$, we have \be\label{defAtoA} \left|{\bar A_{\bar f}\over
A_f}\right|=\left|{\sum_i A_i e^{i(\delta_i-\phi_i)}\over  \sum_i
A_i e^{i(\delta_i+\phi_i)}}\right|. \ee
\subsubsection{CP
violation in the interference of decays with and without mixing}
One can learn CKM phases from decays of neutral $B$ mesons to CP
eigenstates $f_{CP}$.
As a result of $B^0-\bar B^0$ mixing, a state which is $B^0$ at
proper time $t=0$ will evolve into one, denoted $B^0(t)$, which is
a mixture of $B^0$ and $\bar B^0$. Thus there will be one pathway
the final state $f_{CP}$ from $B^0$ throughout the amplitude
$A_{f_{CP}}$ and another from $\bar B^0$ through the amplitude
$\bar A_{f_{CP}}$, which acquires an additional phase $2\beta$
trough the mixing. The CP invariance is violated when the
time-dependent decay rate of $B^0\to f_{CP}$ and that of the CP
conjugated decay $\bar B^0\to f_{CP}$ are different: \be
\Gamma_{B^0\to f_{CP}}(t)\neq \Gamma_{B^0\to f_{CP}}(t)\ee for any
$t$.

The interference of the amplitudes $A_{f_{CP}}$ and $\bar
A_{f_{CP}}$ can differ in the decays $B^0(t)$ and $\bar B^0(t)$
leading to a time dependent asymmetry. To calculate this asymmetry
it is convenient to introduce a complex quantity $\lambda_f$
defined by \be\label{lambdaf} \lambda_{CP}
=\left(\frac{q}{p}\right)_B \frac{\bar A_{f_{CP}}}{
A_{f_{CP}}}=e^{-2 i \beta} \frac{\bar A_{f_{CP}}}{A_{f_{CP}}}. \ee
In the case $\Delta\Gamma_B\ll\Gamma$ one obtains for time
dependent rates
\be\label{widths} \left\{ \begin{array}{c} \Gamma[B^0(t) \to f] \\
\Gamma[\bar B^0 (t) \to f]
\end{array} \right\} \sim e^{- \Gamma t} [ 1 \mp {\cal A}_{CP} \cos \Delta m_B t
 \mp {\cal S}_{CP} \sin \Delta m_B t ]~~~,
\ee where \be\label{AandS}{\cal A}_{CP} \equiv
\frac{|\lambda_{CP}|^2 - 1}{|\lambda_{CP}|^2 + 1}~~,~~~ {\cal
S}_{CP} \equiv \frac{2 {\rm Im} \lambda_{CP}}{|\lambda_{CP}|^2 +
1}~~, \ee where $A\,(\bar A)$ denotes the $B^0\,(\bar B^0)\to
f_{\rm CP}$ decay amplitude. Note that ${\cal S}_f^2 + {\cal
A}_f^2 \le 1$.

The weak phase $\phi_f$ is the phase of $A_f$. Therefore
\be\label{AbarA} \frac{\bar A_{ f}}{A_f}=e^{-2i\phi_f}. \ee Eqs.
(\ref{qpforB}) and (\ref{AbarA}) together imply that for a final
CP eigenstate, \be\label{lamfCP} \lambda_{f_{\rm CP}}=\eta_{f_{\rm
CP}}e^{-2i(\phi_B+\phi_f)}, \ee where $\eta_{f_{\rm CP}}=\pm1$ is
the CP eigenvalue of the final state.

To illustrate the phase structure of decay amplitudes, consider
the process in which the $\bar b$-quark decays through  $\bar b\to
{\it q}\bar {\it q} {\it d}$ decay, where $q$ is an {\it up}-quark
($u$ or $c$), ${\it d}$ is the corresponding {\it down} quark. The
decay Hamiltonian is of the form \be\label{Hdecay} H_d\propto
e^{+i\phi_f}[\bar {\it q} \gamma^\mu(1-\gamma_5){\bar {\it d}}
][\bar b\gamma_\mu(1-\gamma_5){\it q}] +e^{-i\phi_f}[\bar {\it q}
\gamma^\mu(1-\gamma_5)b][\bar {\it d}\gamma_\mu(1-\gamma_5){\it
u}]. \ee

\noindent In this case \be\label{eq:abarovera}\frac{\bar
A_f}{A_f}=\frac{V_{{\it q}b}V_{{\it qd}}^*}{V_{{\it q}b}^*V_{\it
qd}}. \ee We now consider two specific examples.

 \subsection{$B\to J/\psi K_S$. $\sin 2\beta$
measurements} The parameter $\sin2\beta$ is directly accessible
through a study of $CP$ violation in the ``golden decay mode'' of
$B^{0}$ mesons
\begin{equation}
B^{0}~ {\rm or}~\bar B^{0}\rightarrow J/\psi K_{S}.
\label{eq:gold}
\end{equation}
The ``golden'' character of $B^{0}\rightarrow J/\psi K_{S}$
derives from the fact that the final state is a $CP$ eigenstate,
and that this decay mode is dominated by a $CP$ conserving tree
diagram. Any $CP$ violation observed in this mode must, to an
excellent approximation, be attributed to $B^{0}-\bar B^{0}$
mixing. The transitions $B^{0}-\bar B^{0}$ are described, in the
lowest order, by a single  box diagram involving two $W$ bosons
and two $up$-type quarks. The phase of the box diagram was seen to
be $\exp(2\,i\,\beta)$. Therefore measurement of $CP$ violation
effects in this decay mode can be directly interpreted as a
measurement of the $\beta$ angle in the unitarity triangle.

As it was discussed in the preceding subsections, in decays of
neutral $B$ mesons into a CP eigenstate $f_{\rm CP}$, an
observable CP asymmetry can arise from the interference of the
amplitudes for decays with an without $B$--$\bar B$ mixing, i.e.,
from the fact that the amplitudes for $B^0\to f_{\rm CP}$ and
$B^0\to\bar B^0\to f_{\rm CP}$ must be added coherently. For the
time dependent asymmetry  one obtains from Eqs. (\ref{widths}),
(\ref{AandS}) \bea
   A_{\rm CP}(t) & = &
   \frac{\Gamma(\bar B^0(t)\!\to\! f_{\rm CP})
         -\Gamma(B^0(t)\!\to\! f_{\rm CP})}
        {\Gamma(\bar B^0(t)\!\to\! f_{\rm CP})
         +\Gamma(B^0(t)\!\to\! f_{\rm CP})}
= \nonumber\\ &&\frac{2\mbox{Im}(\lambda_{CP})}{1+|\lambda_CP|^2}
   \sin(\Delta m_B t)
   - \frac{1-|\lambda_{CP}|^2}{1+|\lambda_{CP}|^2} \cos(\Delta m_B
   t),
\eea where $\phi_B$ is the introduced above $B$--$\bar B$ mixing
phase (which in the SM equals $-2\beta$).

\begin{figure}
\centerline{\includegraphics[width=0.80\textwidth]{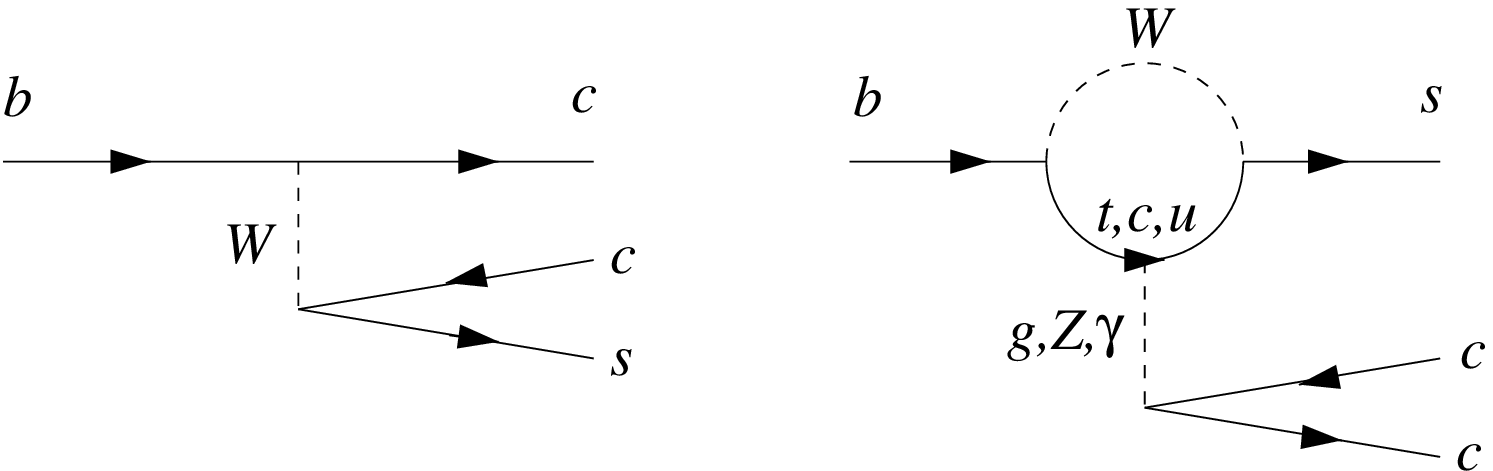}}
\caption{\label{fig:topol}Tree and penguin topologies in $B\to
J/\psi\,K$ decays.}
\end{figure}

The decay is mediated by the quark transition $\bar b\to\bar
cc\bar s$. In the SM, the this decay can proceed via tree level
diagram or via penguin diagrams with intermediate $u$, $c$ and $t$
quarks, as shown in Figure~\ref{fig:topol}. The tree diagram has a
CKM factor $V_{cb}^*V_{cs}$. An amplitude P of penguin diagram has
three terms, corresponding to the three different up-type quarks
inside the loop and can be written in the following schematic form
\be\label{P} P=V_{tb}^*V_{ts} f(m_t)+V_{cb}^*V_{cs}
f(m_c)+V_{ub}^*V_{us} f(m_u),\ee where the $f(m_q)$ is some
function of the quark mass. We use the unitarity relationship to
rewrite the three terms in (\ref{P}) in terms of two independent
CKM factors: \be\label{PR} P=V_{cb}^*V_{cs}[
f(m_c)-f(m_t)]+V_{ub}^*V_{us} [f(m_u)-f(m_t)].\ee The first of
these is the same as that for the tree term, so for the present
discussion it can be considered as part of the ``tree amplitude''.
The second term is suppressed by two factors.   First, ignoring
CKM factors, the penguin graph contribution is expected to be
suppressed compared to the tree graph, because it is a loop graph
and has an additional hard gluon. Second, it is CKM suppressed by
an additional factor of
\be\left|\frac{V_{ub}^*V_{us}}{V_{cb}^*V_{cs}}\right|
\sim\lambda^2.\ee The ``penguin pollution'' to the weak phase is
of order
\begin{equation}\phi_A\sim\lambda^2\left |\frac{P}{T}\right |\sim
1\%,\end{equation} where $|P/T|\sim 0.2$ is the tree-to-penguin
ratio.

Thus we have an amplitude that effectively has only a single CKM
coefficient and hence one overall weak phase, which means there is
no decay-type (direct) $CP$ violation. Indeed, we need at least
two terms with different weak phases to get such an effect. This
then ensures \be \frac{\bar A_{J/\psi K_S}}{A_{J/\psi
K_S}}=\left(\frac{V_{cb}V_{cs}^*}{V_{cb}^*V_{cs}}\right)
\left(\frac{V_{cs}V_{cd}^*}{V_{cs}^*V_{cd}}\right).\ee The last
factor is $(q/p)_K$ in $K^0-\bar K^0$ mixing. This is crucial
because in the absence of $K^0-\bar K^0$ mixing there could be no
interference between ${\bar B}^0\to J/\psi{\bar K}^0$ and $ B^0\to
J/\psi K^0$. Then one finds \be \hat\lambda_{J/\psi
K_S}=-\left(\frac{V_{tb}^*V_{td}}{V_{tb}V_{td^*}}\right)
\left(\frac{V_{cb}V_{cs}^*}{V_{cb}^*V_{cs}}\right)
\left(\frac{V_{cs}V_{cd}^*}{V_{cs}^*V_{cd}}\right)=-\exp(-2i\beta),\ee
where the first factor is the SM value for the $(q/p)_B$ in
$B^0-\bar B^0$ mixing. Recall that for the $B$ meson we expect
$|q/p|_B=1$ to a good approximation. Thus $A_{J/\psi K_S}$
measures Im}$~\lambda_{J/\psi K_S}=\sin 2\beta$: \be
A_{CP}(t)\simeq \sin2\beta\sin(\Delta m_Bt).\ee

The results (\ref{sinbeta}) obtained by Belle experiment and by
Babar experiment are in reasonable agreement among themselves.
Combining Belle and BaBar results with earlier measurements by CDF
at Fermilab $(0.79^{+0.41}_{-0.44})$, ALEPH and OPAL at CERN
$(0.84^{+0.82}_{-1.04}\pm 0.16)$   gives the ``world average''
\begin{equation} \sin 2\beta=0.734\pm 0.054~.
\label{gwa}
\end{equation}

 \subsection{$B\to \pi^+\pi^-$. $\sin 2\alpha$
measurements} To measure the angle $\alpha$, the  most promising
and straightforward approach involves the use of the decay mode
$B^0\to\pi^+\pi^-$ which is an example of the final CP eigenstate.
The interference in this mode between direct decay and the decay
via mixing leads to a CP violating asymmetry as in the charmonium
mode $B^0\to J/\psi K_S.$

However there are several additional complications. The decay
amplitude for $B^0\to\pi^+\pi^-$ contains a contribution from a
tree diagram ($\bar b\to \bar u  ud$) as well as Cabbibo
suppressed penguin diagram which has the flavor structure $\bar
b\to \bar d$ with the final $\bar d d$ pair fragmenting into
$\pi^+\pi^-$. The magnitude of the tree amplitude is $|T|$; its
weak phase is Arg($V^*_{ub}) = \gamma$; by convention its strong
phase is 0. The amplitude of the penguin amplitude  is $|P|$.  The
dominant $t$ contribution in the loop diagram for $\bar b \to \bar
d$ can be integrated out and the unitarity relation $V_{td}
V^*_{tb} = - V_{cd} V^*_{cb} - V_{ud} V^*_{ub}$ used.  The $V_{ud}
V^*_{ub}$ contribution can be absorbed into a redefinition of the
tree amplitude.
By definition, its strong phase is $\delta$. However the remaining
penguin  contributions are not negligible (as in the previous case
$B^0\to J/\psi K_S$) and has a weak phase that is different from
the phase of the tree amplitude, which is zero in the usual
parameterization. Therefore the time dependent asymmetry,
proportional to $\sin\Delta m_B$, which is measured is not equal
to $\sin2\alpha$ but instead will have a large unknown correction.
The presence of the extra contribution also induces an additional
time dependent term proportional to $\cos\Delta m_Bt$.

The time-dependent asymmetries $S_{\pi \pi}$ and $A_{\pi \pi}$
specify both  $\alpha$ and $\delta$, if one has an independent
estimate of $|P/T|$.  One may obtain $|P|$ from $B^+ \to K^0
\pi^+$ using flavor SU(3) \cite{GR95} and $|T|$ from $B \to \pi l
\nu$ using factorization \cite{LR}. An alternative method
\cite{Charles} uses the measured ratio of the $B^+ \to K^0 \pi^+$
and $B^0 \to \pi^+ \pi^-$ branching ratios to constrain $|P/T|$.
The update value of $|P/T|$ is $0.28 \pm 0.06$ \cite{Rosner03}.

\begin{table}
\label{tab:sa} \caption{Values of $S_{\pi \pi}$ and $A_{\pi \pi}$
quoted by BaBar and Belle. The errors for average $S_{\pi \pi}$
and $A_{\pi \pi}$ are taken from \cite{Rosner03}}
\begin{center}
\begin{tabular}{c c c c} \hline
             & BaBar \cite{Bapipi}  & Belle \cite{Bepipi}            &
    Average \\ \hline
$S_{\pi \pi}$ & $0.02\pm0.34\pm0.05$ &
$-1.23\pm0.41^{+0.08}_{-0.07}$ &
    $-0.49 \pm 0.61$ \\
$A_{\pi \pi}$ & $0.30\pm0.25\pm0.04$ & $ 0.77 \pm 0.27 \pm 0.08$ &
    $~~0.51 \pm 0.23$ \\ \hline
\end{tabular}\end{center}
\end{table}

The experimental situation regarding the time dependent
asymmetries is not yet settled. As shown in Table \ref{tab:sa},
BaBar  and Belle  obtain different asymmetries, especially $S_{\pi
\pi}$. At present values of $\alpha~> 90^{\circ}$ are favorated,
but with large uncertainty. It is not yet settled whether ${\cal
A}_{\pi\pi}\neq 0$, corresponding to ``direct'' CP violation.

\subsection{Conclusions}

The significance of the $\sin2\beta$ measurements is that for the
first time a large CP asymmetry has been observed, proving that CP
is not an approximate symmetry of Nature. Rather, the CKM phase
is, very likely, the dominant source of CP violation in low-energy
flavor-changing processes. This opens a new era, in which the
model is expected to be scrutinized through a variety of other $B$
and $B_s$ decay asymmetries. Impressive progress has already been
made in search for asymmetries in several hadronic $B$ decays,
including $B^0 \to \pi^+\pi^-$, $B^{0,\pm} \to K \pi$ and $B^{\pm}
\to DK^{\pm}$. Current measurements are approaching the level of
tightening bounds on the CP-violating phase $\gamma$. These and
forthcoming measurements of $B_s$ decays will enable a cross-check
of the CKM model.

\section{$V_{cb}$ determination}

As discussed in Section 2, $|V_{cb}|$ sets the overall scale for
the lengths of the sides, and $|V_{ub}|$ determines the length of
one side. Precise determinations of both are needed to complement
the measurement of the angles of the unitarity triangle. In this
section we shall discuss exclusive semileptonic decays of
$B$-mesons, in which the $b$-quark decays into a $c$-quark, and
from which one can determine the $|V_{cb}|$ elements of the
CKM-matrix. In principle, $|V_{cb}|$  can be studied in any weak
decay mediated by the W boson. Semileptonic decays offer the
advantage that the leptonic current is calculable and QCD
complications only arise in the hadronic current. Unlike hadronic
decays, there are no final state interactions. One still needs
some understanding of the strong interaction. Some approaches
offer detailed predictions for the QCD dynamics in heavy quark
decays. These predictions allow measurement of $|V_{cb}|$ with
reasonable precision.

\subsection{$\rm B\to D^{\star} \ell \nu$ and $\rm B\to D
\ell \nu$ decays}

\noindent The exclusive $|V_{cb}|$ determination is obtained
studying the $\rm B\to D^{\star} \ell \nu$ and $\rm B\to D
 \ell
\nu$ decays. These decays have been studied in experiments
performed at the $\Upsilon(4S)$ center of mass energy (CLEO
\cite{Cleo_vcb} , Belle \cite{belle-dslnu}) and at the $Z^0$
center of mass energy at LEP (ALEPH \cite{ALEPH_vcb}}, DELPHI
\cite{DELPHI_vcb}, and OPAL \cite{OPAL_vcb}).

\subsubsection{Kinematics}

The hadronic form factors for semileptonic  decays are defined as
the   Lorentz--invariant   functions  arising  in  the covariant
decomposition   of  matrix  elements  of  the  vector and  axial
currents. It is conventional to parameterize these matrix elements
by a set of scalar form factors. The most appropriate to the
heavy-quark limit is the set of form factors $h_i(w)$, which are
defined separately for the vector and axial currents:
\be\label{formfactors}
  \langle D(v')|\,\bar c\gamma^\mu b\,|B(v)\rangle =
  h_+(w)(v+v')^\mu+h_-(w)(v-v')^\mu,
  \ee
  \be
\langle D(v')|\,\bar c\gamma^\mu\gamma^5 b\,|B(v)\rangle
  = 0,\ee
\be
  \langle D^*(v',\epsilon)|\,\bar c\gamma^\mu b\,|B(v)\rangle
  =
  h_V(w)i\varepsilon^{\mu\nu\alpha\beta}
  \epsilon_\nu^*v'_\alpha v_\beta,
\ee \be
  \langle D(v')|\,\bar c\gamma^\mu\gamma^5 b\,|B(v)\rangle
   =
  h_{A_1}(w)(w+1)\epsilon^{*\mu}-\epsilon^*\cdot v
  [h_{A_2}(w)v^\mu+h_{A_3}(w)v^{\prime\mu}]\,
\ee  where meson  states are denoted as $|P(v)>$ for a
pseudoscalar state and $|V(v,\varepsilon>)$ for a vector state,
where $v$ is the 4-velocity of a state and $\varepsilon$ is the
polarization vector, $w=v\cdot v'$
is the velocity transfer which is linearly related to $q^2$, the
invariant mass of the $W$. Other linear combinations of form
factors are also used in the literature, see {\it e.g.}
\cite{Odonnell1997}

In the case of  heavy--to--heavy  transitions,  in the  limit  in
which the active quarks have infinite  mass, all the form factors
are given in terms of a single function ${\cal F}(w)$, the
Isgur--Wise form factor \cite{IW}:\bea\label{ffrelations}
  &&h_+(w)=h_V(w)=h_{A_1}(w)=h_{A_3}(w)={\cal F}(w),\nonumber\\
  &&h_-(w)=h_{A_2}(w)=0.
\eea In the realistic  case of finite quark masses these relations
are modified:  each form factor  depends  separately on the
dynamics of the process.

\subsubsection{The decay $\rm B\to D^{*} \ell \nu$ in Heavy Quark Effective Theory}

Heavy Quark Effective Theory (HQET), see {\it e.g.} \cite{MW00},
predicts that the differential partial decay width for
$B\rightarrow D^{\star}\ell \nu$, $d\Gamma/dw$, is related to
$V_{cb}$ through:
\begin{equation}
\label{btod*} \frac{d\Gamma}{dw}(B\rightarrow D^{\star}\ell \nu) =
\frac{G_F^2 |V_{cb}|^2}{48\pi^3}{\cal K}(w){\cal F}_{D^*}^2(w),
\end{equation}
where
${\cal K}(w)$ is a known phase space factor:
\begin{equation}
{\cal K}(w)=(m_B{-}m_{D^*})^2 m_{D^*}^3
    \sqrt{w^2{-}1}\,(w{+}1)^2
\left[ 1 + \frac{4w}{w{+}1}
     \frac{m_B^2-2w m_Bm_{D^*}+m_{D^*}^2}{(m_B-m_{D^*})^2}
    \right].
\end{equation}
The function ${\cal F}_{D^*}(w)$ is the form factor for the $B$ to
$D^*$ transition, {\it i.e.}, the Isgur-Weise function combined
with perturbative and power corrections. The precision with which
$V_{cb}$ can be extracted is limited by the theoretical
uncertainties in the evaluation of these corrections.

In the infinite quark mass limit in the kinematical point where
$D^*$ is at rest in the $B$ rest frame the wave function overlap
is 1, {\it i.e.}, ${\cal F}_{D^*}=1$. There are several different
corrections to the infinite mass value ${\cal F}_{D^*}(1)=1$:
\begin{equation}
{\cal F}_{D^*}(1) =\eta _{QED}\eta _A \left[ 1 + \delta _{1/m_Q^2}
+ ...\right],~~~Q=c,b
\end{equation}
The correction ${\cal O}(1/m_Q)$ vanishes by virtue of Luke's
theorem \cite{Luke90}. $QED$ corrections  $\eta_{QED}\approx
1.007$ up to leading logarithms. $QCD$ radiative corrections to
two loops give $\eta _A = 0.960\pm 0.007$. Different estimates of
the $1/m_Q^2$ corrections yield \be 1 + \delta _{1/m_Q^2} =\ 0.91
\pm 0.04\ee

The analytical expression of ${\cal F}_{D^*}(w)$ is not known
a-priori, and this introduces an additional uncertainty in the
determination of ${\cal F}_{D^*}(1)|V_{cb}|$. In an experiment one
measures the decay rate as function of $w$ and extrapolates to
$w=1$. As the kinematically allowed range of $w$ is small ($w\in
[1.0,1.5]$), the form factor is approximated as a Taylor expansion
around $w=1$. \be\label{rho}{\cal F}_{D^*}(w) = {\cal F}_{D^*}(1)
(1 + (w-1)\rho^2 + c(w-1)^2). \ee Fig. \ref{btodstar} shows the
latest CLEO measurement \cite{Cleo_vcb} of ${\cal
F}_{D^*}|V_{cb}|$ as a function of $w$. The results of the fits of
the latest experiments are given in Tabl. \ref{table:bd}.
Averaging the data
one gets
\begin{equation}{\cal F}_{D^*}(1) |V_{cb}| = (38.3  \pm 1.0) \times
10^{-3}.\end{equation} This gives the most updated value quoted
from \cite{AB02}
\begin{equation}
\label{vcbd*}
 |V_{cb}|=(42.1 \pm 1.1_{exp} \pm 1.9_{theo})\times 10^{-3}.
\end{equation}
\begin{table}
\begin{center}
\begin{tabular}{|c|c|c|}

 \hline Experiment   & $|V_{cb}|\times 10^{3}$ &
$\rho^2$
\\ \hline
CLEO  \cite{Cleo_vcb}      &  43.3$\pm$  1.3$\pm$ 1.8 & 1.61$\pm$ 0.09$\pm$ 0.21  \\
 Belle  \cite{belle-dslnu}  &  36.0$\pm$  1.9$\pm$ 1.8 & 1.45$\pm$ 0.16$\pm$ 0.20  \\
ALEPH \cite{ALEPH_vcb}      &  33.8$\pm$  2.1$\pm$ 1.6 & 0.74$\pm$ 0.25$\pm$ 0.41  \\
DELPHI  \cite{DELPHI_vcb}   &  36.1$\pm$  1.4$\pm$ 2.5 & 1.42$\pm$ 0.14$\pm$ 0.37  \\
 OPAL    \cite{OPAL_vcb}    &  38.5$\pm$  0.9$\pm$ 1.8 & 1.35$\pm$ 0.12$\pm$ 0.31  \\
\hline

\end{tabular}
\end{center}
\caption{\label{table:bd} Various experimental results for
$|V_{cb}|$. For details see \cite{AB02}. $\rho^2$ is the slope of
the form factor at zero recoil as defined in (\ref{rho}).}
\end{table}

\subsection{$B\to D\ell\nu$} The decay $B\to D\ell\nu$ can be
analyzed in the same way as $B\to D^*\ell\nu$ decay. The
differential decay width for $B\to D\ell\nu$ decay is
\begin{equation}
{d\Gamma\over dw}= {G_F^2 |V_{cb}|^2 \over 48 \pi^3} (M_B +M_D)^2
M_D^3 (w^2-1)^{3/2} {\cal F}_D^2(w) \label{e:kme:dgdw_dlnu}
\end{equation}
where different form factor ${\cal F}_D(w)$ is assumed.  The
precision with which $|V_{cb}|$ can be determined is not as good
as for $B\to D^*\ell\nu$ because of smaller branching fraction,
larger backgrounds and an additional kinematic suppression factor
$w^2-1$ (compare Eqs. (\ref{btod*}) and (\ref{e:kme:dgdw_dlnu})).
Nonetheless it provides complementary information and provides a
test of HQET predictions for the relationships between the form
factors for semileptonic decays $B\to D$ and $B\to D^*$.

Theoretical predictions for ${\cal F}_D(1) $ are: $1.03\pm 0.07$
(quark model \cite{SI95}) and $0.98\pm0.7$ (QCD sum rules
\cite{LNN94}). A quenched lattice calculation gives ${\cal
F}(1)=1.058^{+0.020}_{-0.017}$ \cite{H99}. Using ${\cal
F}_D(1)=1.0\pm 0.07$ PDG 2002 quotes the value
\begin{equation}
 |V_{cb}|=(41.3 \pm 4.0_{exp} \pm 2.9_{theo})\times 10^{-3},
\end{equation}
consistent with (\ref{vcbd*}).
\begin{figure}
\centerline{\includegraphics[height=80mm,keepaspectratio=true]{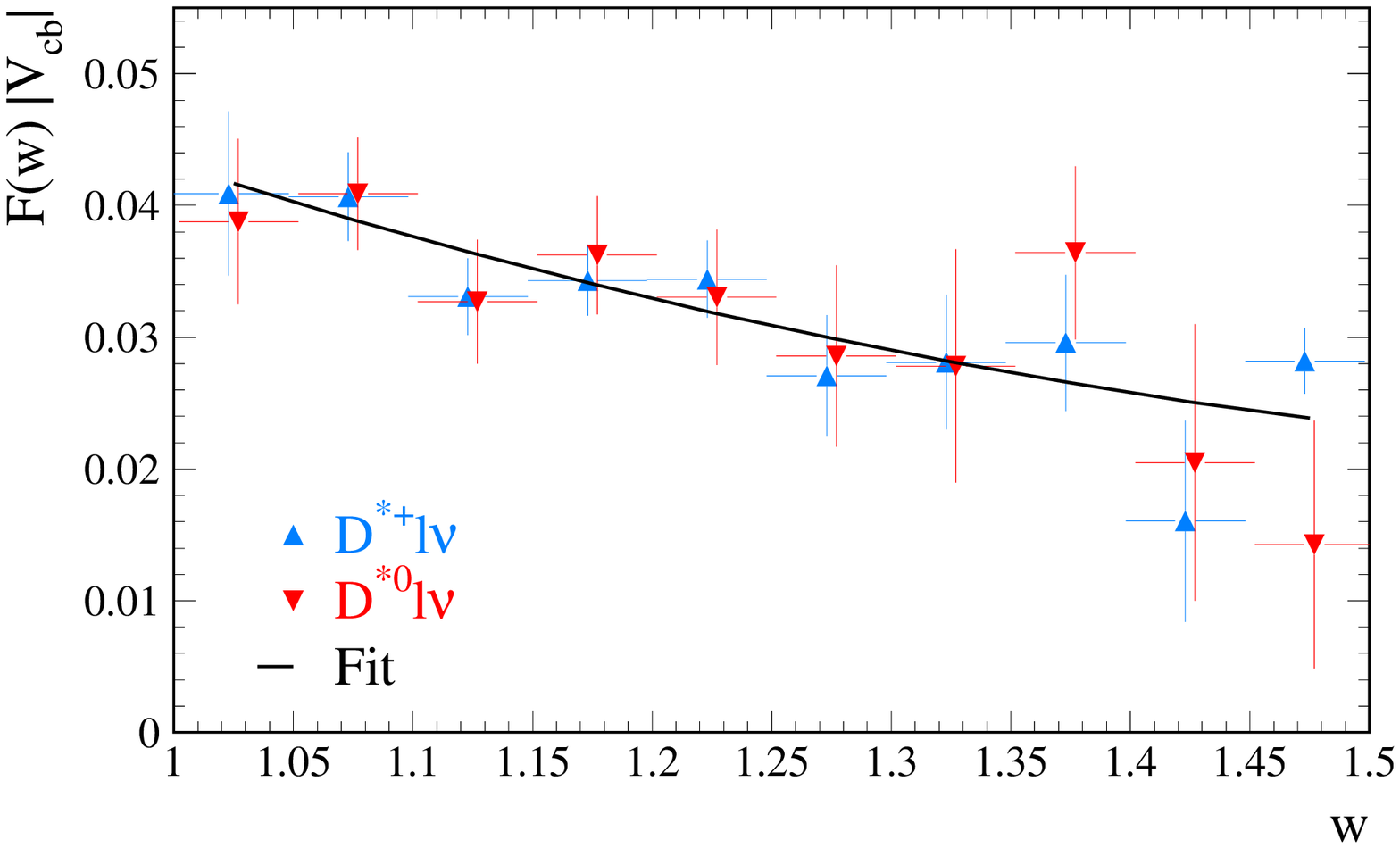}}
\caption{\label{btodstar}Overlay of ${\cal F}_{D^*}(w)|V_{cb}|$
where the points are $B\to D^*\ell\nu$ data}
\end{figure}

\subsection{Inclusive semileptonic decays}
Alternatively, $|V_{cb}|$ can be extracted  from measuring of
electron energy spectra in  inclusive semileptonic  $B \to X_c
\ell \nu $ decay . Inclusive measurements are employed to avoid
the need for form factors, relying on HQET for the necessary quark
level input.

The measurement \cite{Aubert2002} employs the method introduced by
ARGUS \cite{Argus93} and later used by CLEO \cite{CLEO}, in which
$B\bar B$ events are tagged by the presence of a high momentum
lepton. As a tag, electrons are chosed with center-of-mass frame
momentum $1.4 GeV/c~<p^*<~2.3 GeV/c$. A second electron in the
event is taken as the signal lepton for which the condition $p^*
> p_{min}$ is required to avoid large backgrounds at lower momenta. Signal
electrons are mostly from primary B decays if they are accompanied
by a tag electron of opposite charge (unlike-sign). Those with a
tag of the same charge (like-sign) originate predominantly from
secondary decays of charm particles produced in the decay of the
other B meson. Inversion of this charge correlation due to
$B^0\bar B^0$ mixing is treated explicitly, and unlike-sign pairs
with both electrons originating from the same B meson are isolated
kinematically. With a small model-dependence on the estimated
fraction of primary electrons below $p^*=0.6 $ GeV/c, the
semileptonic B branching fraction is inferred from the background
corrected ratio of unlike-sign electron pairs to tag electrons.

The $B\to X_ce\nu$ electron spectrum measurement \cite{CLEO} shown
in Fig.~\ref{lepspec3} is an observed spectrum above 0.6 GeV. In
events with a high momentum lepton tag and an additional electron,
the primary electrons ($b\to c \ell^- X$) are separated from
secondary electrons from charm decays ($b\to c X; c \to \ell^+ Y$
) using angular and charge correlations.

\begin{figure}
\centerline{\includegraphics[height=80mm,keepaspectratio=true]{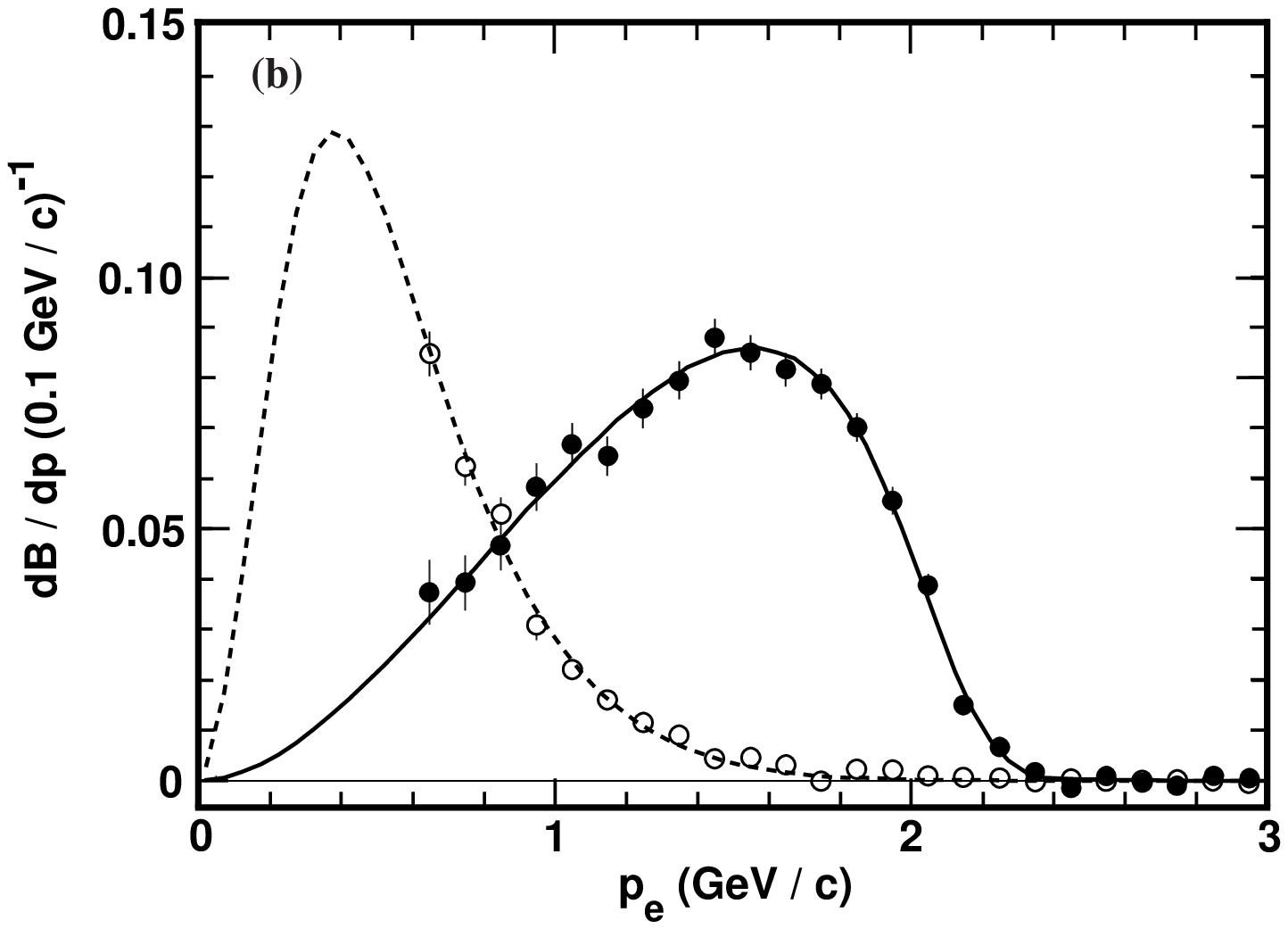}}
\caption{\label{lepspec3}Electron momentum spectrum from $B\to X_
e\nu$ (solid circles) and $b\to c \to Y e\nu$ (open circles)
\cite{{CLEO}}. The curves show the best fit to the ISGW model with
23\% $B\to D^{**}\ell\nu$.}
\end{figure}

For beauty hadrons with $m_b\gg\Lambda_{QCD}$ or $\bar\Lambda$, a
QCD related scale of order 400 MeV (see below), one can use an
operator product expansion (OPE) \cite{BSU97} combined with HQET.
The spectator model decay rate is the leading term in a OPE
expansion controlled by the parameter $\bar\Lambda/m_b$.
Non-perturbative corrections to the leading approximation arise
only to order $1/m_b^2$. The key issue in this approach is the
ability to separate non-perturbative corrections, which can be
expressed as a series in powers of $1/m_b$, and perturbative
corrections, expressed in powers of $\alpha_s$. Quark-hadron
duality is an important {\it ab initio} assumption in these
calculations \cite{bigiduality}. An unknown correction may be
associated with this assumption. Arguments supporting a possible
sizeable source of errors related to the assumption of
quark-hadron duality have been proposed \cite{nathan}. This issue
needs to be resolved with further measurements.

The OPE result for inclusive decay width $\Gamma_{SL}$ reads
\begin{eqnarray}
\label{gamma_inclusive} \Gamma_{SL}& = & \frac{G_F^2 m_b^5}{192
\pi^3}\cdot
\left(1-a_1\frac{\alpha_s}{\pi}-a_2(\frac{\alpha_s}{\pi})^2
+...\right)\nonumber\\&&\times\left((1+\frac{\lambda_1}{2m_b^2})f(\rho)
+\frac{\lambda_2}{2m_b^2}g(\rho)~+~...\right),
\end{eqnarray}
where $a_1=1.54,~a_2=1.43\beta_0$ ($\beta_0$ being the $\beta$
function)
 are coefficients of the perturbative expansion, $m_b(\mu)$ and $m_c(\mu)$ are short scale quark
masses (in particular, $\rm m_b(\mu\sim 1~ \rm GeV) = 4.58 \pm
0.09\ GeV$), $f(\rho)$ and $g(\rho)$ are known parton phase space
factors,
 \begin{equation}
f(\rho)=1-8\rho+8\rho^3-\rho^4-12\rho^2\log~\rho,\end{equation}\begin{equation}
g(\rho)=-9+24\rho-72\rho^2+72\rho^3-15\rho^4-36\rho^2\log~\rho,\end{equation}

\noindent with $\rho=m_c^2/m_b^2$.

The parameters $\lambda_1$ and $\lambda_2$ are matrix elements of
the HQET expansion, which have the following intuitive
interpretations: $\lambda_1$ is proportional to the kinetic energy
of the $b$-quark in the $B$ meson and $\lambda_2$ is the energy of
the hyperfine interaction of the $b$-quark spin and the light
degrees of freedom in the meson.  The third HQET parameter,
$\bar\Lambda$, representing the energy of the light degrees of
freedom is introduced to relate the $b$-quark and $B$ meson
masses, through the expression:
\begin{equation}
m_b = {\bar{m}_B}-{\bar{\Lambda}}+\frac{{\lambda}_{1}}{2 m_b},
\end{equation}
where ${\bar{m}_B}$ is the spin-averaged  mass of $B$ and $B^*$
(${\bar{m}_B} = 5.313$ GeV$/c^2$). A similar relationship holds
between the $c$-quark mass $m_c$ and the spin-averaged charm meson
mass (${\bar{m}_D} = 1.975$ GeV$/c^2$).

The parameter $\lambda _2$ can be extracted from the $\rm B^*-B$
mass splitting and found to be
\begin{equation}
\lambda _2= 0.128\pm 0.010~ {\rm GeV}^2 ,\end{equation} whereas
the other parameters need more elaborate measurements. The aim of
the new inclusive studies is to determine $\lambda_1$ and $\bar
\Lambda$ from experiment and thereby decrease the theoretical
uncertainty which comes when extracting $|V_{cb}|$ from
$\Gamma_{SL}$.

The first stage of this experimental program has been completed
recently. The CLEO collaboration has measured the shape of the
photon spectrum in $\rm b\rightarrow s \gamma$ inclusive decays
\cite{Chen01}. Its first moment (sensitive to $\bar\Lambda$),
giving the average energy of the $\gamma$ emitted in this
transition, is related to the $b$ quark mass. This corresponds to
the measurement of the parameter
\begin{equation}\overline{\Lambda}=+0.35\pm 0.07\pm 0.10~~GeV\end{equation}

For semileptonic decays $B\to X_c\ell\nu$, two methods to
determine $\bar\Lambda$ and $\lambda_1$ are known. The first
method measures the first and second hadronic mass moments while
the second method uses
the measured shape of the lepton ($\ell=e,\mu$) energy spectrum to
determine $\bar\Lambda$ and $\lambda_1$, trough its energy
moments, which are also predicted by HQET. The truncated moments
with a lepton momentum cut $p_{\ell}= 1.5$ GeV
\begin{equation}
{\mbox{R}}_{0} =\frac{\int_{1.7}^{} (d \Gamma_{SL}/dE)
dE_l}{\int_{1.5}^{} (d \Gamma_{SL}/dE) dE}, \label{r2}
\end{equation}
and
\begin{equation}
{\mbox{R}}_{1} =\frac{\int_{1.5}^{} E_l (d \Gamma_{SL}/dE)
dE}{\int_{1.5}^{} (d {\Gamma_{SL}} /dE) dE}. \label{r1}
\end{equation}

\noindent are employed to decrease sensitivity of the measurement
to the secondary leptons from the cascade decays $b\to
c/d\ell\nu$. The theoretical expressions for these moments
\cite{kme:gremm}  are evaluated by integrating over the lepton
energy in the decay $b\rightarrow c \ell \bar{\nu}$ for the
dominant $\Gamma _c$ component. Constraints on $\bar\Lambda$ and
$\lambda_1$ obtained from the CLEO measurements of $R_1$ and $R_2$
are shown in Fig. \ref{parameters}. They correspond to \be
\bar\lambda= 0.39\pm 0.03_{stat}\pm 0.06_{syst}\pm
0.12_{theo},~~~\lambda_1=-0.25\pm 0.02_{stat}\pm 0.05_{syst}\pm
0.14_{theo}.\ee

\begin{figure}
\centerline{\includegraphics[height=90mm,keepaspectratio=true]{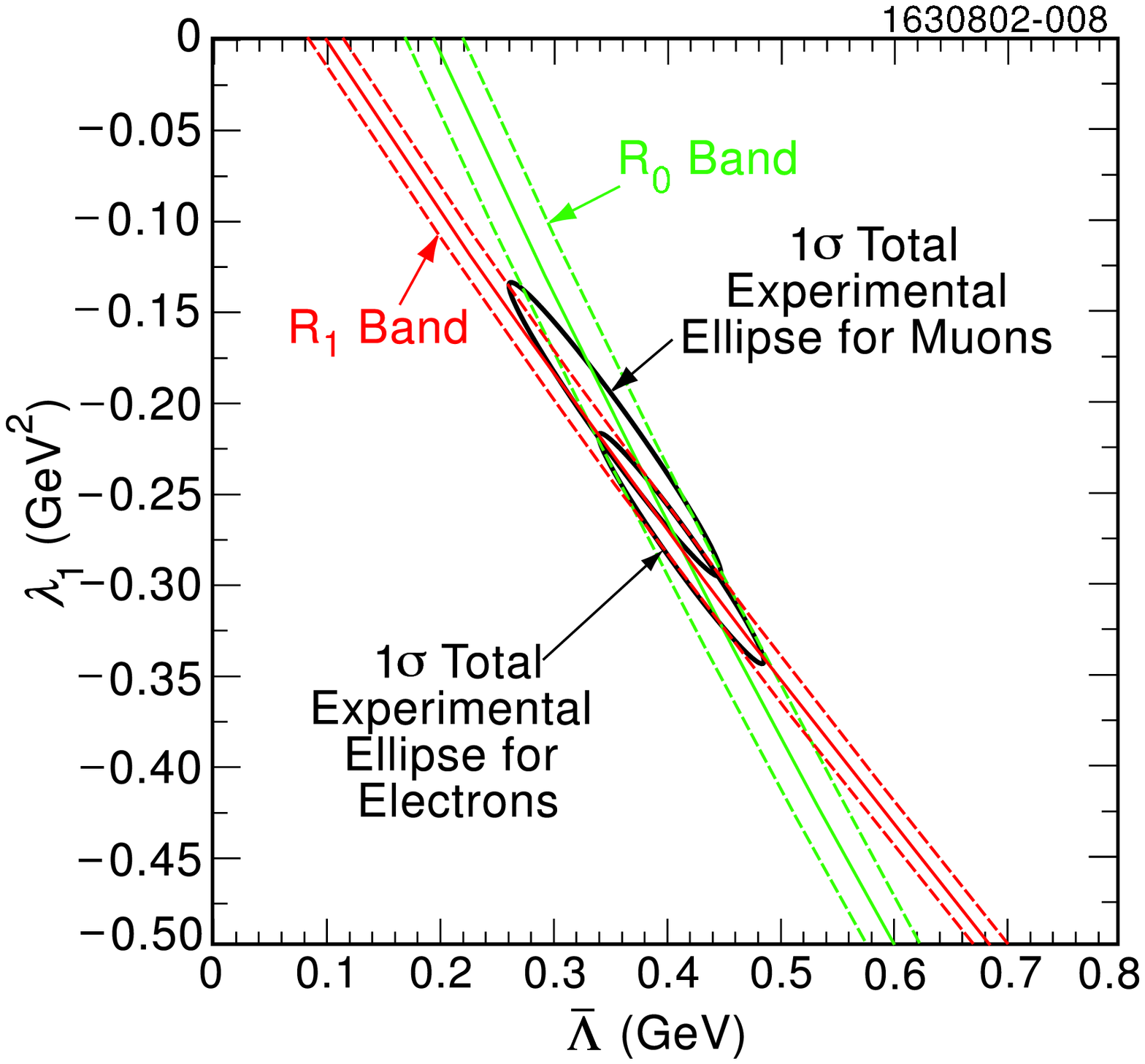}}
\caption{\label{parameters}Constrains on the HQET parameters
$\lambda_1$ and $\bar\Lambda$ from  measurements of the moments
R$_0$ and R$_1$.  The contours represent $\Delta\chi^2=1$ for the
combined statistical and systematic errors on the measured
values.}
\end{figure}

Using the expression of the full semileptonic decay width given in
Eq.~(\ref{gamma_inclusive}) and the experimental value
$\Gamma_{SL}^{exp}=(0.43 \pm 0.01){\times} 10^{-10}$ MeV
\cite{AB02}, one can extract $|{\mbox{V}}_{cb}|$:

\begin{equation}
|{\mbox{V}}_{cb}|=(40.8\pm 0.5|_{\Gamma_{SL}}\pm0.4|_{\lambda_1,
\bar{\Lambda}}\pm 0.9|_{th})\times 10^{-3},
\end{equation}
where the first uncertainty is from the experimental value of the
semileptonic width, the second uncertainty is from the HQET
parameters (${\lambda}_{1}$ and $\bar{\Lambda}$), and the third
uncertainty is the theoretical one. Non-quantified uncertainties
are associated with a possible quark-hadron duality violation.

\subsection{$|V_{ub}|$}
The methods which are currently available for probing $V_{ub}$ are
unfortunately plagued by dependence on phenomenological models
whose uncertainties are difficult to quantify reliably.  As a
result, despite many experimental efforts present constraints on
this parameter are unacceptably weak.  The analyses which have
been used fall into two classes, inclusive decays of the form
$B\to X_ul\nu$, and exclusive transitions such as
$B\to(\pi,\rho)l\nu$.

The inclusive decay rate has the advantage that it can be
predicted in the form of a systematic expansion in powers of
$1/m_b$. Experimentally measurements of $b\to u\ell\nu$ which are
sensitive to $|V_{ub}|$ are difficult due to the overwhelming
background from the Cabibbo favored $b\to c\ell\nu$ decays. At
present, this background can be suppressed only by confining
oneself to kinematic regions in which only charmless final states
can contribute, such as $E>2.2\,$GeV or $M(X)<1.9\,$GeV.
Unfortunately, the OPE techniques which allow one to calculate
reliably the total inclusive rate breaks down when the phase space
is restricted in this way. Phenomenological models must then be
used to reconstruct the rate in the unobserved kinematic regions,
and the model independence of the analysis is lost.

The inclusive analysis includes a wide kinematic range to avoid
losing signal statistics but then pays the price of quark-hadron
duality and a fine-tuned modeling of charm backgrounds.  Combining
the observed yield of lrptons in the end-point momentum interval
2.2-2.6~GeV/c with the recent data on $B\to X_s\gamma$ and using
HQET CLEO \cite{vubincleo} report the value of
\be\label{vubin}|V_{ub}|=(4.08\pm0.34\pm0.44\pm0.16\pm0.24)\times
10^{-3},\ee where the first two uncertainties are experimental and
the last two are from the theory.

 Exclusive transitions are easier
to study experimentally. On the other hand theoretical predictions
of exclusive decay channels are polluted by the ignorance of the
physics of quark hadronization\footnote{Various approaches to this
problem have been proposed (for example, heavy quark symmetry,
chiral expansions in the soft pion limit, dispersion relations,
QCD sum rules, and lattice calculations), but in many of these
cases significant model dependence remains.}. CLEO restricts
itself to exclusive final states ($B\to\pi l\nu$, $\rho l\nu$)
using $\nu$-reconstruction, in which there is a more favorable
signal-to-noise but a considerable uncertainty in the
form-factors. The CLEO exclusive result \be |V_{ub}|=
(3.25\pm0.30\pm0.55)\times 10^{-3},\ee in which the first error is
experimental and the second is theoretical, is consistent with
that obtained from inclusive measurements.

\subsection{Conclusions}

At present our knowledge of $\lambda_1$ and $\bar\Lambda$ limits
the precision we can achieve on from inclusive semileptonic $B$
decays. The aim of the new inclusive analyses is to determine
$\lambda_1$ and $\bar \Lambda$ from experiment and thereby
decrease the theoretical uncertainty which comes when extracting
$|V_{cb}|$ from $\Gamma_{SL}$.  Each analysis alone provides two
constraints, allowing a measurement of $\bar\Lambda$ and
$\lambda_1$. Combining the two analyses over-constrains the theory
parameters thus allowing a test of the theoretical framework and
experimental understanding of $b$-quark decays.

While experimental errors have reached $1-2\%$ level, the dominant
uncertainties remain of theoretical origin. High precision tests
of HQET, checks on possible violations of quark-hadron duality in
semileptonic decays. Experimental determination of $m_b$,
$m_b-m_c$ and $\lambda_1$ are needed to complete this challenging
experimental program.

\section{A bit of phenomenology. Electron spectra in semileptonic $B$ and $B_c$ decays.}

Electron energy spectrum in inclusive $B\to X_c\ell\nu$ decays can
be also treated using OPE. The result (away from the endpoint of
the spectrum) is that the inclusive differential decay width
$d\Gamma/dE$ may be expanded in $\Lambda/m_b$. The leading term
(zeroth order in $\Lambda/m_b$) is the free quark decay spectrum,
the subleading term vanishes, and the subsubleading term involves
parameters from the heavy quark theory, but should be rather
small, as it is  of order $(\bar\Lambda/m_b)^2$.

However, the calculation of the lepton energy spectrum in OPE
shows the appearance of singular distributions
$\delta^{(n)}(E-m_b/2) $ near the end point where $E=m_b/2$. The
non adequacy  of the approach is also evident from the fact that
although $m_b/2$ is the largest lepton energy available for a free
quark decay, the physical endpoint corresponds to $E=m_B/2$. In
this {\it windows} bound state effects, due to the Fermi motion of
the heavy quark, become important and  the $1/m_b$ expansion has
to be replaced by an expansion in twist. To describe this region
one has to introduce a so--called ``shape function''
\cite{Bigi94}, \cite{Neubert94} which in principle introduces a
hadronic uncertainty. This is quite analogous to what happens for
the structure function in deep--inelastic scattering in the region
where the Bjorken variable $x_B\to 1$. A model independent
determination of the shape function is not available at the
present time, therefore a certain model dependence in this region
seems to be unavoidable, unless lattice data become reasonably
precise.

Two phenomenological approaches had been applied to describe
strong interaction effects in the inclusive weak decays: the
parton ACM model \cite{ACM82} amended to include the motion of the
heavy quark inside the decaying hadron, and the ``exclusive
model'' based on the summation of different channels, one by one
\cite{ISGW89}.

The various light--front (LF) approaches to consideration of the
inclusive semileptonic transitions were suggested in Refs.
\cite{JP94}--\cite{KNST99}. In Refs. \cite{JP94}, \cite{MTM96} the
Infinite Momentum Frame prescription  $p_b=x p_B$, and,
correspondingly, the floating $b$ quark mass $m_b^2(\xi)=
x^2m_B^2$ have been used. The transverse $b$ quark momenta were
consequently neglected.
In Ref. \cite{KNST99} the $b$--quark was considered as an
on--mass--shell particle with the definite mass $m_b$ and
the effects arising from the $b$--quark transverse motion in the
$\bar B$--meson were included. The corresponding ans\"atz of Ref.
\cite{KNST99} reduces to a specific choice of the primordial LF
distribution function $|\psi(\xi,p^2_{\bot})|^2$, which represents
the probability to find the $b$ quark carrying a LF fraction $\xi$
and a transverse momentum squared $p^2_{\bot}=|{\bf p}_{\bot}|^2$.
As a result, a new parton--like formula for the inclusive
semileptonic $b\to c,u$ width has been derived \cite{KNST99},
which is similar to the one obtained by Bjorken {\it et al.}
\cite{BDT92} but in case of infinitely heavy $b$ and $c$ quarks.

\subsection{ACM model}
The ACM model was originally developed to consider in detail the
endpoint of the lepton spectrum in order to estimate a systematic
error in modeling  the full spectrum. It incorporates some of the
corrections related to the fact that the decaying $b$ quark is not
free, but in a bound state. It was explicitly constructed to avoid
mention of a $b$ quark mass. The model is extensively used in the
analysis of the lepton energy spectrum in semi-leptonic decays. It
reproduces very well numerically the shape of the semi-leptonic
spectra  at least in its regular part.

  The model treats the $B$ meson
with the mass $m_B$ as consisting of the heavy $b$ quark plus a
spectator with fixed mass $m_{sp}$; the latter usually represents
a fit parameter. The spectator quark has a momentum distribution
$\Phi ({\bf p}^2)$ (${\bf p}$ is its three-dimensional momentum).
The momentum distribution is usually taken to be Gaussian:
 normalized so that
\begin{equation} \int\Phi({\bf p}^2)p^2dp=1.\end{equation}
The decay spectrum  is determined by the kinematics constrains on
the $b$
 quark.
The energy-momentum conservation in the $B$ meson vertex implies
that the $b$ quark energy is \be E_b=m_B-\sqrt{{\bf
p}^2+m_{sp}^2},\ee  thus the $b$ quark cannot possess a definite
mass. Instead, one obtains a ``floating'' $b$ quark mass
\begin{equation}
\label{fm} (m_b^f)^2=m_B^2+m_{sp}^2-2m_B\sqrt{{\bf p}^2+m_{sp}^2},
\end{equation} which depends on $|{\bf p}|^2$. The lepton spectrum is first
obtained from the spectrum $d\Gamma_b^{(0)}(m_f,E)/dE$ of the $b$
quark of invariant mass $m_b^f$ (in the $b$ quark rest frame)
\begin{equation} \label{22} {d \Gamma_b^{(0)}(m_b,E) \over d
E}={G_F^2 m_b^4 \over 48\pi^3}{x^2 (x_{max}-x)^2 \over (1-x)^3}
\left[(1-x)(3-2x)+(1-x_{max})(3-x)\right] \end{equation}  with
$x={2 E_e/m_b}$, $x_{max}= 1-\rho$, and $\rho={m_c^2/m_b^2}$, then
boosting back to the rest frame of the $B$ meson and averaging
over the weight function $\Phi ({\bf p}^2)$ .
\begin{equation}
\label{3} {d \Gamma_B \over dE}=\int\limits_0^{p_{max}} dp p^2
\Phi({\bf p}^2){(m_b^f)^2 \over 2p E_b} \int_{E_-}^{E_+} {dE'
\over E'} {d \Gamma^{(0)}_b(m_f,E') \over dE'}~. \end{equation}
 The perturbative
corrections are neglected for the moment. In Eq. (\ref{3})
\begin{equation}
p_{max}=\frac{m_B}{2}-\frac{m_c^2}{2m_B-4E}~, ~~~E_{\pm
}=\frac{Em_b^f}{E_b\mp |{\bf p}|}~. \end{equation} In fact the
upper limit of integration in (\ref{3}) is not $E^+$ but ${\it
min}(E_+,E_{max})$, where \begin{equation}
E_{max}=\frac{m_B-m_{sp}}{2}\left(1-\frac{m_c^2}{(m_B-m_{sp})^2}\right).
\end{equation} These expressions conclude the kinematical
analysis in the ACM model.

\subsection{B-meson on the Light-Front}

The elegance and simplicity of the Light-Front (LF) approach
results from the analogy uf relativistic field theories quantized
in the LF to non-relativistic quantum mechanics. In fact this
correspondence runs deep and there is exact isomorphism between
the Galilean subgroup of the Poincar\'e group and the symmetry
group of  two dimensional quantum mechanics. LF theory also
provides a support for the intuitive quark-parton picture of bound
states in QCD. The purpose of this subsection is to illustrate the
attractive feature of the Lf approach in the simplest fashion by
working out a concrete example of inclusive semileptonic $B$
decays \cite{KNST99}. Other applications can be found in
\cite{Odonnell1997},\cite{DGNS97}.

Similar to the ACM model the LF quark model treats the beauty
meson as consisting of the heavy $b$ quark plus a spectator quark.
Both quarks have fixed masses, $m_b$ and $m_{sp}$, though. This is
at variance with the ACM model, that has been introduced in order
to avoid the notion of the heavy quark mass at all. The
calculation of the distribution over lepton energy in the LF
approach does not requires  any boosting procedure but is based on
the standard Lorentz--invariant kinematical analysis.

There are three independent kinematical variables in the inclusive
phenomenology: the lepton energy $E_{\ell}, q^2$, where
$q=p_{\ell}+p_{\nu_{\ell}}$, and the invariant mass
$M^2_X=(p_B-q)^2$ of a hadronic state. Introducing the
dimensionless variables $y=2E_{\ell}/m_B$, $t=q^2/m^2_B$, and
$s=M^2_X/m^2_B$, the differential decay rate for semileptonic $B$
decay can be written as
\begin{eqnarray}\label{13} \frac{d\Gamma_{SL}}{dy}&=&
\frac{G_F^2m_B^5}{64\pi^3}|V_{cb}|^2 \int\limits^{t_{
max}}_{0}dt\int\limits^{s_{max}}_{s_0}ds \\
&&\times\left\{tW_1+\frac{1}{2}[y(1+t-s)-y^2-t]W_2+
t[\frac{1+t-s}{2}-y]W_3+\ldots\right\},\nonumber \end{eqnarray}
where the structure functions $W_i=W_i(s,t)$ appear in the
decomposition of the hadronic tensor $W_{\alpha\beta}$ in Lorentz
covariants. The ellipsis in (\ref{13}) denote the terms
proportional to the lepton mass squared. The kinematical limits of
integration can be found from equation \begin{equation} \label{14}
\frac{s}{1-y}+\frac{t}{y}\leq 1 \end{equation} They are given by
$0\leq y\leq (1-\rho)$, where $\rho=m_c^2/m_b^2$,
$s_{max}=1+t-(y+t/y)$, and $t_{max}=y[1-\rho/(1-y)]$.

In a parton model LF inclusive semileptonic $ B_c \to
X_{Q'}\ell\nu_{\ell} $ decay is treated in a direct analogy to
deep-inelastic scattering. An approach is
 based on the hypothesis of quark--hadron duality.
This hypothesis assumes that the inclusive decay probability for
which no reference to a particular hadronic state is needed equals
to one into the free quarks. The basic ingredient is the
expression for the hadronic tensor $W_{\alpha\beta}$ which is
given through the optical theorem by the imaginary part of the
quark box diagram describing the forward scattering amplitude:
\begin{equation}\label{16} W_{\alpha\beta}=\int
L^{(cb)}_{\alpha\beta}(p_c,p_b)\delta
[(p_b-q)^2-m_c^2]\frac{|\psi(\xi,p^2_{\bot})|^2}{\xi}\theta(\varepsilon_c)
d\xi d^2p_{\bot}, \end{equation} where a quark tensor $
L^{(cb)}_{\alpha\beta} (p_c,p_b) $ is defined as

\begin{equation} \label{17}
    L^{cb}_{\alpha\beta}(p_c,p_b)  = \frac{1}{4} \sum_{spins}
    \bar{u}_{c} O_{\alpha} u_b \cdot \bar{u}_b O^+_{\beta}u_{c}
 =2(p_{c\alpha}p_{b\beta}+p_{c\beta}p_
{b\alpha}-g_{\alpha\beta}(p_cp_b)+i\epsilon_ {\alpha\beta
\gamma\delta}p^{c\gamma}p^{b\delta}) \end{equation} and the factor
$1/\xi$ in Eq. (\ref{16}) comes from the normalization of the $B$
meson vertex \cite{DGNS97}.

\noindent Eq. (\ref{16}) amounts to averaging the perturbative
decay distribution over motion of heavy quark governed by the
distribution function $|\psi(\xi,p^2_{\bot})|^2$. In this respect
the approach is similar to the parton model in deep inelastic
scattering, although it is not really a parton model in its
standard definition. The normalization condition reads
\begin{equation} \label{18} \pi\int\limits_0^1 d\xi\int
dp^2_{\bot}|\psi(\xi,p^2_{\bot})|^2=1. \end{equation} The function
$\theta(\varepsilon_c)$ where $\varepsilon_c$ is the $c$--quark
energy is inserted in Eq. (\ref{16}) for consistency with the use
of valence LF wave function to calculate the $b$--quark
distribution in the $B$--meson.

Recall that the endpoint for the quark decay spectrum is
\begin{equation}
y_{max}^b=(m_b/m_B)(1-m_c^2/m_b^2),\end{equation} whereas the
physical endpoint is \begin{equation}
y_{max}=1-m^2_D/m_B^2.\end{equation} where $m_D$ is the $D$ meson
mass. The endpoint for the LF electron spectrum is in fact not
$y_{max}$ but \begin{equation}
y_{max}^{LF}=1-m^2_c/m_B^2.\end{equation}
 This is the direct
consequence of the $p^2_{\bot}$ integration in Eq. (\ref{23})
\cite{KNST99}. Note that $y_{max}^{LF}$ coincides with
$y_{max}^{ACM}$ with accuracy $\sim m_{sp}/m_B$. For $m_c\sim 1.5$
GeV the difference between $y_{max}^{LF}$ and $y_{max}$ is of the
order $10^{-2}$.

\subsection{The distribution function of the $b$ quark}
An explicit representation for the $B$-meson Fock expansion in QCD
is not known. {\it A priory}, there is no connection between
equal--time (ET) wave function $w({\bf k}^2)$ of a constituent
quark model and LF wave function $\psi(x,p^2_{\bot})$. The former
depends on the center--of--mass momentum squared $k^2=|{\bf
k}|^2$, while the latter depends on the LF variables $\xi$ and
$p_{\bot}^2$. However, there is a simple operational connection
between ET and LF wave functions \cite{C92}. This is model
dependent enterprise but has its close equivalent in studies of
electron spectra using the ACM model. The idea is to find a
mapping between the variables of the wave functions   that will
turn a normalized solution $w({\bf k}^2)$ of the ET equation of
motion into a normalized solution $\psi(\xi,p^2_{\bot})$ of the
different looking LF equation of motion. That will allows us to
convert the ET wave function, and all the labor behind it, into a
usable LF wave function. This procedure amounts to a series of
reasonable (but naive) guesses about what the solution of a
relativistic theory involving confining interactions might look
like.

Specifically, one converts from ET to LF momenta by leaving the
transverse momenta unchanged, ${\bf k}_{\bot}={\bf p}_{\bot}$ and
letting
\begin{equation}
\label{19}
p_{iz}=\frac{1}{2}(p_i^+-p_i^-)=\frac{1}{2}(p_i^+-\frac{p^2_{i{\bot}}
+m^2_i}{p_i^+}) \end{equation} for both the $b$--quark $(i=b)$ and
the quark--spectator $(i=sp)$. Here $p_i^{\pm}=p_{i0}\pm p_{iz}$
with $\sum p_i^+=p_B^+=m_B$ (in the B meson rest frame).

In what follows we identify $\Phi({\bf k}^2)=|w({\bf k}^2)|^2$
with the Gaussian distribution
\begin{equation} \label{gauss} \Phi({\bf k}^2)={4 \over \sqrt{\pi}
p_F^3}\exp\left(-\frac{{\bf k}^2}{ p_F^2}\right).
\end{equation}

The simple calculation yields  \be \label{21}
|\psi(\xi,p^2_{\bot})|^2=
\frac{4}{\sqrt{\pi}p^3_f}\exp\left(-\frac{
p^2_{\bot}+p^2_z}{p^2_f}\right)\left|\frac{\partial
p_z}{\partial\xi}\right|, \ee where  \be \label{p_z}
p^2_z(\xi,p^2_{\bot})=\frac{1}{2}\left((1-\xi)m_B-\frac{p^2_{\bot}
+m^2_{{\rm sp}}}{(1-\xi)m_B}\right),~~~\xi=\xi_{{\rm sp}},\ee and
\be \left|\frac{\partial p_z}{\partial
\xi}\right|=\frac{1}{2}\left(m_B+\frac{p^2_{\bot} +m^2_{{\rm
sp}}}{(1-\xi)^2m_B}\right). \ee

The calculation of the structure functions $W_i(t,s)$ in the LF
parton approximation (\ref{16}) is straightforward. The result is

\begin{equation} \label{WLF} W_i(t,s)=\int w_i(s,t,\xi)\delta
[(p_b-q)^2-m_c^2]\frac{|\psi(\xi,p^2_{\bot})|^2}{\xi}\theta(\varepsilon_c)
d\xi d^2p_{\bot}, \end{equation} where the structure functions
$w_i(s,t,\xi)$ are defined in the same way as  $W_i(s,t)$ in
(\ref{13}) but for the free quark decay. Explicit expressions for
$W_i(s,t)$  can be obtained  using Eq. (\ref{17}), they are given
in Ref. \cite{KNST99}. Eq. (\ref{WLF}) differs from the
corresponding expressions of Refs. \cite{JP94} and \cite{MTM96} by
the non--trivial dependence on $p^2_{\bot}$ which enters both
$|\psi(\xi,p^2_{\bot})|^2$ and argument of the $\delta$--function.
For further details see \cite{KNST99}.

\subsection{The choice of $m_b$}

An important technical issue that appear in the problem is the
definition of the quark mass $m_b$. The semileptonic decay rate is
proportional to $m_b^5$, thus any uncertainty in the definition of
heavy quark mass transfers into a huge uncertainty in the
predicted rate. The problem is to find a definition consistent
with that of HQET.

In the ACM model, it is known \cite{RS93}, \cite{BSUV94} that once
the semileptonic width $\Gamma_{{\rm ACM}}$ is expressed in terms
$m_b^{ACM}~=~<m_b^f>$ (that is nothing but the floating mass
$m_b^f({\bf p}^2)$ of Eq. (\ref{fm}) averaged over the
distribution $\Phi({\bf p}^2$) ), the correction to first order in
$1/m_b$ both to the inclusive semileptonic width and to the
regular part of the lepton spectrum can be absorbed into the
definition of the quark mass, in full agreement with the general
HQET  statement of the absence of the $1/m_b$ correction in total
width.

The choice of $m_b$ in the LF approach was first addressed in the
context of the LF model for $b\to s\gamma$ transitions
\cite{KKNS99}. It was shown that the LF model can be made agree
with HQET provided $m_b^{LF}$ is defined from the requirement of
the vanishing of the first moment of the distribution function.
This condition coincides with that used in HQET to define the pole
mass of the $b$--quark. In this way one avoids an otherwise large
(and model dependent) correction of order $1/m_b$ but at expense
of introducing the shift in the constituent quark mass which
largely compensates the bound state effects. It has been also
demonstrated that the values of $m_b^{LF}$ found by this procedure
agree well with the average values $<m_b^f>$ in the ACM model.
Accepting the identification $m_b^{LF}=m_b^{ACM}$, the similar
agreement but for the semileptonic $b\to c$ decays has been found
in Ref. \cite{GKN01}.

\subsection{Electron energy spectra. LF model vs ACM model}

In Table \ref{table:acmvslf} for various values of $p_F$,  the
values of the total semileptonic width for the free quark with the
mass $m_b=<m_b^f>$ and the  $B$ meson semileptonic widths,
calculated using the LF and ACM approaches, respectively, are
given. In the last two columns, shown are the fractional deviation
$\delta=\Delta \Gamma_{SL}/\Gamma_{SL}^b$ (in per cent) between
the semileptonic widths determined in the LF and ACM models and
that of the free quark. The agreement between the LF and ACM
approaches  for integrated rates is excellent for small $p_F$.
This agreement is seen to be breaking down at $p_F\geq 0.4$ GeV,
but even for $p_F\sim 0.5$ GeV the difference between the ACM and
LF inclusive widths is still small and is of the order of a per
cent level.
\begin{table}
\begin{center}
\begin{tabular}{|c|c|c|c|c|c|c|}
\hline\hline $p_F$& $<m_b^f>$ & $\Gamma_{SL}^b$ &
$\Gamma_{SL}^{ACM}$ & $\Gamma_{SL}^{LF}$ & $\delta^{ACM}$ &
$\delta^{LF}$ \\ \hline\hline
0.1 & 5.089 & 0.1007 & 0.1005 & 0.1005 & 0.2 & 0.2\\
0.2 & 5.004 & 0.0906 & 0.0902 & 0.0901 & 0.4 & 0.5\\
0.3 & 4.905 & 0.0799 & 0.0792 & 0.0789 & 0.9 & 1.2\\
0.4 & 4.800 & 0.0696 & 0.0688 & 0.0682 & 1.1 & 2.0\\
0.5 & 4.692 & 0.0602 & 0.0592 & 0.0584 & 1.7 & 3.0\\
\hline
\end{tabular}
\end{center}
\caption{\label{table:acmvslf} Comparison of the LF and ACM
results for the semileptonic integrated rates. In all cases
$m_{sp}=0.15$ GeV and $m_c=1.5$ GeV and the radiative corrections
are neglected. A momentum distribution of the b-quark is taken in
the standard Gaussian form (\ref{gauss}) with the Fermi momentum
$p_F$. $|V_{cb}|=0.04$.}
\end{table}

\begin{figure}
\centerline{\includegraphics[height=180mm,keepaspectratio=true]{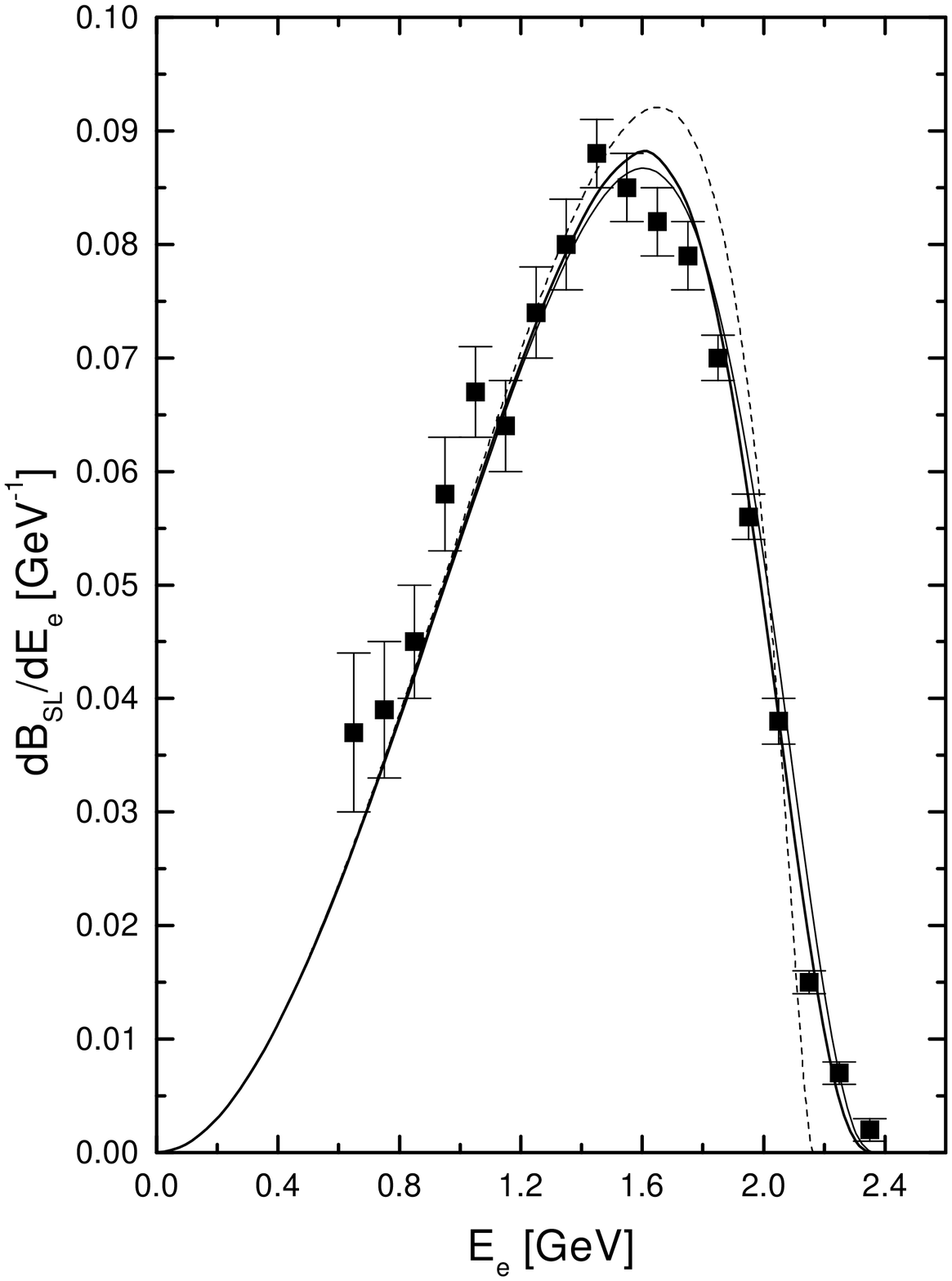}}
\caption{\label{b_spectra}The predicted electron energy spectrum
compared with the CLEO data \cite{CLEO}. The calculation uses
$p_f=0.4$ GeV, $m_b=4.8$ GeV, $m_c=1.5$ GeV, and $\alpha_s=0.25$
for the perturbative corrections. Thick solid line is the LF
result, thin solid line is the ACM result, dashed line refers to
the free quark decay. The spectra normalized to $10.16\%$,
$10.23\%$, and $10.36\%$, respectively. $|V_{cb}|=0.042$.}
\end{figure}

Fig. \ref{b_spectra} shows the three theoretical curves for
electron spectrum in inclusive $B\to X_c\ell\nu_{\ell}$ decays are
presented for the LF, ACM and free quark models.  This is a direct
calculation of the spectrum and not a $\chi^2$ fit. A more
detailed fit can impose constrains on the distribution function
and the mass of the charm quark. Such the fit should also account
for detector resolution. The overall normalization of the electron
energy  spectra is
\begin{equation}
BR_{LF}=10.16\%,~~~BR_{ACM}=10.23\%~~~BR_{free}=10.37\%,
\end{equation}
in agreement with the experimental finding \cite{CLEO}
$BR_{SL}=(10.49\pm 0.17\pm 0.43) \%$.

The calculations implicitly include the ${\it O}(\alpha_s)$
perturbative corrections arising from gluon Bremsstrahllung and
one--loop effects which modify an electron energy spectra at the
partonic level (see {\it e.g.} \cite{JK89} and references
therein). It is customary to define a correction function $G(x)$
to the electron spectrum $d\Gamma_b^{(0)}$ calculated in the tree
approximation for the free quark decay through \begin{equation}
\label{23} \frac{d\Gamma_b}{dx}=
\frac{d\Gamma^{(0)}_b}{dx}\left(1-\frac{2\alpha_s}{3\pi}G(x)\right),
\end{equation} where $x=2E/m_b$. The function $G(x)$ contains the
logarithmic singularities $\sim\ln^2(1-x)$ which for $m_c=0$
appear at the quark-level endpoints $x_{max}=1$. This singular
behaviour at the end point is clearly a signal of the inadequacy
of the perturbative expansion in this region. The problem  is
solved by taking into account the bound state effects \cite{JP94}.
Since the radiative corrections must be convoluted with the
distribution function the endpoints of the perturbative spectra
are extended from the quark level to the hadron level and the
logarithmic singularities are eliminated.
\begin{figure}
\centerline{\includegraphics[height=180mm,keepaspectratio=true]{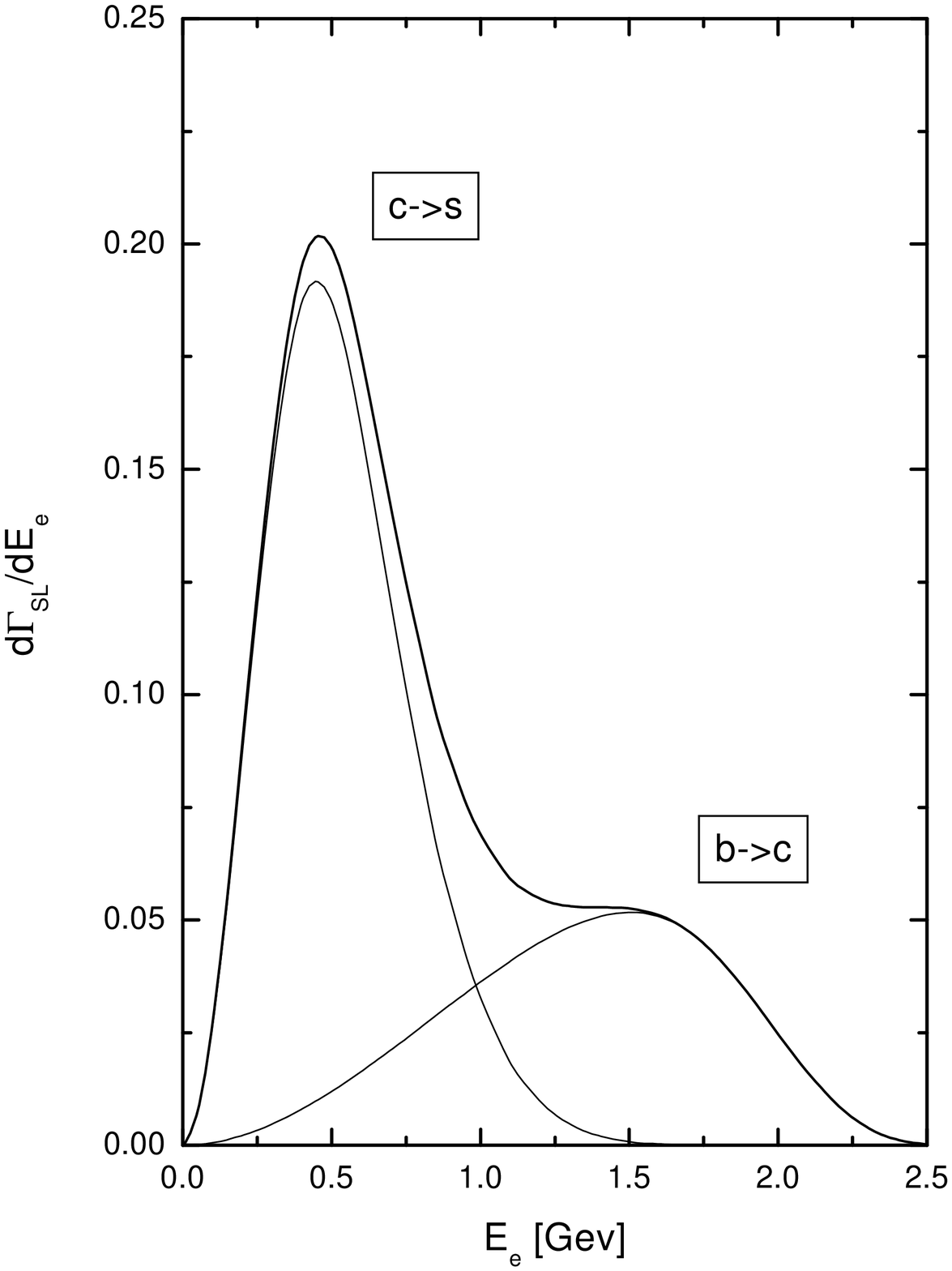}}
\caption{\label{bc_92}The predicted electron energy spectrum in
semileptonic $B_c$ decays.  The calculation uses $p_F=0.92$ GeV,
$m_b=5.0$ GeV, $m_c=1.5$ GeV, and $\alpha_s=0.25$ for the
perturbative corrections.}
\end{figure}

\subsection{$B_c$ decays} The semileptonic decay of $B_c$ consists
of two contributions, $\Gamma_{SL}=\Gamma_{SL}^b+\Gamma_{SL}^c$.
which are, respectively. $b\to \bar cW^+$ with $c$-quark as
spectator and $\bar c\to\bar sW^-$ with $b$ quark as spectator.
Since these processes lead to different the final states, their
amplitudes do not interfere. In the simplest view, $b$ and $\bar
c$ are free, and the total semileptonic width is just the sum of
the $b$ and $\bar c$ semileptonic widths, with $c$-decay
dominating. Approximating this by
$\Gamma_{SL}(B_c)=\Gamma_{SL}(B)+\Gamma_{SL}(D)$ yields
$\Gamma_{SL}(B_c)\sim 0.22$ ps$^{-1}$. This estimate is modified
by strong interaction effects.

Fig. \ref{bc_92} shows the lepton energy spectrum in the decay
$B_c\to X e\nu_e$. This calculation refers to the case $m_{b}=5$
GeV, $m_c=1.5$ GeV as chosen in Ref. \cite{BB96}. The free quark
semileptonic widts are $\Gamma_{SL}^{c,free}=0.218$ ps$^{-1}$, and
$\Gamma_{SL}^{b,free}=0.090$, ps$^{-1}$. The Fermi momentum $p_F$
is chosen as $p_F=0.92$ GeV corresponding to the Isgur-Scora
Model.  Like the OPE formalism the LF  approach leads to a
reduction of the free quark decay rates caused by binding,
$\Gamma_{SL}^{b,bound}=0.090$, ps$^{-1}$
$\Gamma_{SL}(B_c)=0.18$~ps$^{-1}$, but the bound state corrections
for $c\to s$ semileptonic rate are substantially larger than those
reported in \cite{BB96}. The result of Ref. \cite{BB96} would
correspond to a very soft $B_c$ wave function with $p_F\sim 0.5$
GeV, which is seemed to be excluded by existing constituent quark
models.

\begin{figure}
\centerline{\includegraphics[height=180mm,keepaspectratio=true]{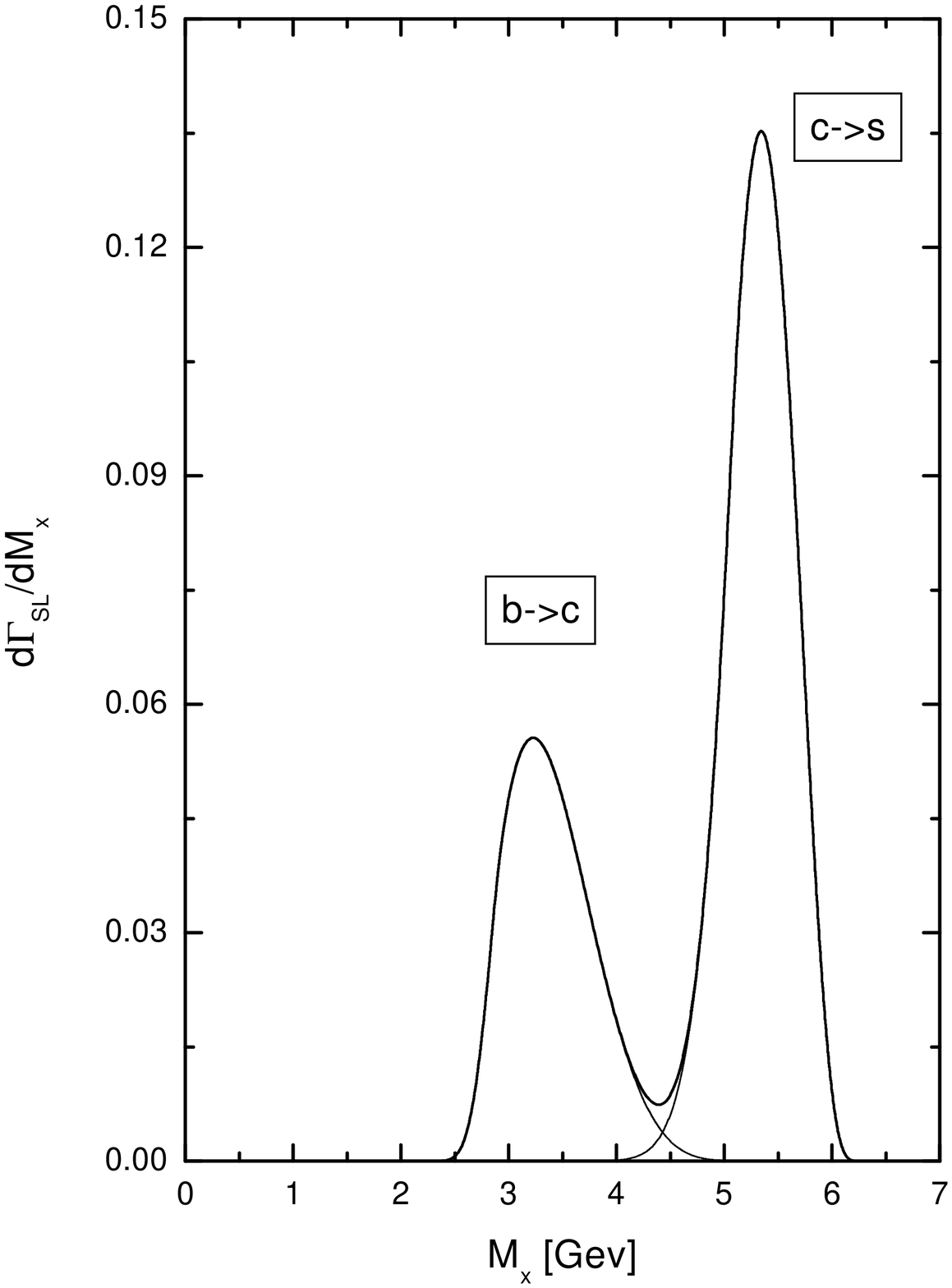}}
\caption{\label{mx_92}The predicted hadronic mass distribution
spectrum in semileptonic $B_c$ decays. The calculation uses
$p_F=0.92$ GeV, $m_b=5.0$ GeV, $m_c=1.5$ GeV}
\end{figure}
Finally, we note that the theoretical results for the electron
spectrum can be translated into predictions for the hadronic mass
spectrum. In Fig. \ref{mx_92} we show the invariant mass
distribution of the hadrons recoiling against $\ell\nu$. The LF
predictions for hadronic mass spectra must be understood in the
cense of quark--hadron duality. The true hadronic mass spectrum
may have resonance structure that looks rather different from
inclusive predictions. Inclusive calculations predict a continuum
which is given by the inclusive spectrum and is dual to a large
number of overlapping resonances.

\section*{Acknowledgement}  I am grateful to K.Boreskov for careful reading of
the manuscript
and valuable suggestions. This work was supported by NATO
grant $\#$ PST.CLG.978710, RFBR grant $\#$ 03-02-17345 and PRF
grant for leading scientific schools $\#$ 1774.2003.2.

\renewcommand{\refname}{References}


\begin{thebibliography}{99}
\bibitem{CKM} Cabibbo~N, Phys. Rev. Lett. {\bf 10}, 531 (1963);
Kobayashi~M. and Maskawa~T., Prog. Theo. Phys. {\bf 49}, 652
(1973)
\bibitem{Buras03} Buras~A., CP violation in B and K decays, Lectures
given at the 41 Schladming School in Theoretical Physics,
Schladming, February 22-28, 2003, arXiv: hep-ph/0307293
\bibitem {JIM70} Glashow~S,, Iliopoulos~J., Maiani~L., Phys. Rev.
{\bf D2},  1285 (1970)
\bibitem{Argus1997} Albrecht, H,. {\it et al.}, Argus collaboration,
Phys. Lett., {\bf B192}, 245 (1997)
\bibitem{BaBar} Aubert~B. {\it et al.} [BaBar Collaboration], Improved
measurement
of the CP-violating asymmetry amplitude sin2beta, arXiv:
hep-ex/0203007
\bibitem{Belle} Abe~K. {\it et al.} [Belle Colaboration], Phys. Rev.
{\bf D66} 032007 (2002)
\bibitem{Q01} Quinn~E., {\it B} physics and {\it CP} violation,
Lectures given at Particle Physics School, ICTP, Trieste, July
2001, arXiv: hep-ph/0111177
\bibitem{nyr01} Nir~Y., CP violation: A New
Era,  Lectures given at the 55th Scottish University Summer School
in Physics, St.Andrews, Scottland, August 7-20, 2001, arXiv:
hep-ph/0109090 \bibitem{Neubert01} Neubert~M., arXiv:
hep-ph/0110304; Nir~Y., arXiv: hep-ph/0208080
\bibitem{AB02} Artuso~M. and Barberio~E., ref. \cite{pdg}
[arXiv: hep-ph/0205163]
\bibitem{Altarelli} Altarelli~G., Feruglio~F., to be
published in {\it Proceedings of the X International Workshop on
Neutrino Telescopes}, March 11-14 2003, Venezia, Italy, arXiv:
hep-ph/0306265
\bibitem{pdg} Hagiwara~K. {\it et al.}, ({\it Particle Data Group}), Phys.
Rev. {\bf D 66}, 010001 (2002)
\bibitem{Jarlskog} Jarlskog~C., Phys. Rev. Lett. {\bf 55}, 1039 (1985)
\bibitem{CLEObsg}
Ahmed~S. {\it et al.}  [CLEO Collaboration], arXiv:
hep-ex/9908022.
\bibitem{ALEPHbsg}
Barate~R. {\it et al.} [ALEPH Collaboration], Phys. Lett. {\bf
B429}, 169 (1998).
\bibitem{greubhurthQCD}
Greub~C. and Hurth~T., Nucl. Phys. (Proc. Suppl.) {\bf B74}, 247
(1999) [arXiv: hep-ph/9809468].
\bibitem{wolf-prl}Wolfenstein~L., Phys. Rev. Lett. {\bf 51}, 1945 (1983).
\bibitem{BLO94} Buras~A.J., Lautenbacher~M.E. and Ostermaier~G.,
Phys. Rev. {\bf D50}, 3433 (1994)  [arXiv: hep-ph/9403384]
\bibitem{Buras_Fleischer_HeavyFlavorsII} Buras~A.J. and Fleishner~R,
in ``Heavy Flavors II,'' eds. Buras~A.J. and Linder~M., World
Scientific, Singapore 1998
\bibitem{Hocker:2001xe}
H\"ocker~A., Lacker~H., Laplace~S., and Le Diberder~F., ``A new
approach to a global fit of the CKM matrix,'' Eur. Phys. J. C {\bf
21},  225 (2001)
\bibitem{V03} Vysotsky~M., arXiv: hep-ph/0307218
\bibitem{S02} Schneider~O., ref. \cite{pdg} [arXiv
:hep-ph/0206171]
\bibitem{GR95} Gronau~M and Rosner~J.L., Phys. Rev. {\bf D53}, 2516 (1996);
Phys. Rev. Lett., {\bf 76} 1200 (1996).
\bibitem{LR} Luo~Z.and Rosner~J.L., Phys. Rev. {\bf D65} 054027 (2002).
\bibitem{Charles} Charles~J., Phys. Rev, {\bf D59}, 054007 (1999).
\bibitem{Rosner03} Rosner~J.L., arXiv: hep-ph/0305315
\bibitem{Bapipi} BaBar collaboration, Aubert.~B.~{\it et al.}, Phys. Rev. Lett. {\bf 89}, 281802(2002).
\bibitem{Bepipi} Belle collaboration, Abe, K. {\it et al}, Phys. Rev., {\bf D68}, 012001 (2003).
\bibitem{Cleo_vcb}
Briere~R.A. {\it et al.}, CLEO Collaboration,
Phys.Rev.Lett. {\bf 89} 081803 (2002), arXiv: hep-ex/0203032.
\bibitem{belle-dslnu}Abe~K. {\it et al.} (Belle collaboration),
Phys. Lett, {\bf B526}, 247 (2002) [arXiv: hep-ex/0111060].
\bibitem{ALEPH_vcb}Buskulic~D.{\it et al.} (ALEPH collaboration),
Phys. Lett. {\bf B335},  373 (1997)
\bibitem{DELPHI_vcb}Abreu~P.  {\it et al.} (DELPHI collaboration),
Phys. Lett. {\bf B 510}, 55 (2001).
\bibitem{OPAL_vcb}Abbiendi~G.  {\it et al.} (OPAL collaboration),
Phys. Lett. {\bf B 482}, 15 (2000)
\bibitem{Odonnell1997} Demchuk~N.B., Kulikov~P.Yu, Narodetskii~I.M.,
O'Donnell~P.J., Phys. Atom. Nucl. {\bf 60} 1292 (1997); Yad. Fiz.
{\bf 60}, 1429 (1997)
\bibitem{IW}  Isgur~N. and Wise~M.B., Phys. Lett. {\bf
B232}, 113 (1989); Phys. Lett. {\bf B237}, 527 (1990).
\bibitem{MW00} Neubert~M. and Wise~M.B., Heavy-Quark Physics,
Cambridge University Press, Cambridge 2000
\bibitem{Luke90} Luke~M., Phys. Lett. {\bf B 252}, 447 (1990)
\bibitem{SI95} Scora~D. and Isgur~N.,
Phys. Rev. {\bf D52}, 2783 (1995).
\bibitem{LNN94} Ligeti~Z., Nir~Y. and Neubert~M.,
Phys. Rev. {\bf D49}, 1302 (1994).
\bibitem{H99} Hashimoto~S. {\it et al.}, Phys. Rev. {\bf
D61}, 014502 (1999)
\bibitem{Aubert2002} Aubert~B. {\it et al.} CLEO collaboration, Phys.Rev.
{\bf D67}, 031101 (2002) [arXiv: hep-ex/0208018]
\bibitem{Argus93} Albrecht~H. {\it et al.} ARGUS collaboration,
Phys. Lett. {\bf B318}, 377 (1993)
\bibitem{CLEO} Barish~B. {\em et al.}, CLEO collaboration, Phys. Rev. Lett. {\bf 76}, 1570
(1995)
\bibitem{BSU97} Bigi~I., Shifman~M., and Uraltsev~N.G.,
Annu. Rev. Nuc. Part. Sci., {\bf 47}, 591 (1997).
\bibitem{bigiduality}Bigi~I. and  Uraltsev~N.G.,
Int.J.Mod.Phys. {\bf A16}, 5201 (2001)
\bibitem{nathan}Isgur~N., Phys. Lett. {\bf B448}, 111 (1999)
\bibitem{vubincleo} Bomheim~A.{\it et al.)} (CLEO collaboration),
Phys. Rev. Lett. {\bf 88}, 231803-1 (2002)
\bibitem{Chen01} Chen~S. {\it et al.}, CLEO collaboration,  Phys.Rev.Lett. {\bf 87}
251807 (2001) [arXiv: hep-ex/0108032]
\bibitem{kme:gremm} Gremm~M. and Kapustin~A., Phys. Rev. {\bf D55}, 6924 (1997)
[arXiv: hep-ph/9603448].
\bibitem{Brier02} Briere~R.A. {\it et al.}, CLEO collaboration,
arXiv: hep-ex/0209024
\bibitem{Bigi94} Bigi~I. {\it et al.}, Int. J. Mod. Phys. {\bf A9}, 2467
(1994)
\bibitem{Neubert94} Neubert~M., Phys. Rev. {\bf D49}, 3392,
4623 (1994)
\bibitem{ACM82} Altarelli~G., Cabibbo~N., Corbo~G., Maiani~L., and Martinelli~G.,
Nucl. Phys. {\bf B202}, 512 (1982)
\bibitem{ISGW89}  Isgur~N. {\it et al.}, Phys. Rev.
{\bf D39}, 799 (1989).
\bibitem{JP94} Jin~C.H., Palmer~M.F., and Paschos~E.A., Phys. Lett. {\bf
B329}, 364 (1994)
\bibitem{MTM96} Morgunov~V.L., Ter-Martirosyan~K.A., Phys. Atom. Nucl. {\bf
59}, 1221 (1996)
\bibitem{GNST97} Grach~I.L., Narodetskii~I.M., Simula~S.,
and Ter-Martirosyan~K.A., Nucl. Phys. {\bf B592}, 227 (1997)
\bibitem{KNST99} Kotkovsky~S., Narodetskii~I.M., Simula~S.,
and Ter-Martirosyan~K.A., Phys. Rev. {\bf D60}, 114024 (1999)
\bibitem{BDT92} Bjorken~J., Dunietz~I., and Taron~M., Nucl. Phys. {\bf A371},  111 (1992)
\bibitem{DGNS97} Demchuck~N.B., Grach~I.L., Narodetskii~I.M., and Simula~S.,
Phys. At. Nucl. {\bf 59}, 2152 (1996)
\bibitem{C92} Coester~F., Prog. Part. Nucl. Phys. {\bf 29}, 1 (1992)
\bibitem{RS93} Randall~L. and Sundrum~R., Phys. Lett. {\bf B312}, 148 (1993)
\bibitem{BSUV94} Bigi~I., Shifman~M., Uraltsev~N.G., Vainstein~A., Phys. Lett. {\bf
B328}, 431 (1994)
\bibitem{KKNS99} Keum~Y.-Y., Kulikov~P.Yu, Narodetskii~I.M., Song~H.S., Phys. Lett.
{\bf B471}, 72 (1999)
\bibitem{GKN01} Grach~I.L., Kulikov~P.Yu, Narodetskii~I.M., JETP
Letters, {\bf 73}, 317 (2001)
\bibitem{JK89} Jezabek~M. and K\"uhn~J.H., Nucl. Phys. {\bf B320}, 20 (1989)
\bibitem{DKNO03} Datta~A., Kulikov~P.Yu., Narodetskii~I.M. and O'Donnell~P.J., in preparation
\bibitem{BB96} Beneke~M, Buchalla~G., Phys. Rev. {\bf D53}, 4991 (1996)
\end{thebibliography}
\end{document}